\documentclass[12pt, preprint]{aastex}

\slugcomment{Submitted to {\it The Astrophysical Journal}}

\shorttitle{External Compton and Internal Shock Model}
\shortauthors{Joshi, Marscher \& B\"ottcher}

\begin{document}

\title{Seed Photon Fields of Blazars in the Internal Shock Scenario}

\author{M. Joshi\altaffilmark{1}, A. P. Marscher\altaffilmark{1}, and 
  M. B\"ottcher\altaffilmark{2, 3}}
\affil{Institute for Astrophysical Research, Boston University,
\\725 Commonwealth Ave., Boston, MA 02215, USA}
\affil{Centre for Space Research, North-West University,
\\Potchefstroom Campus, Potchefstroom 2520, South Africa}
\affil{Astrophysical Institute, Department of Physics and Astronomy,
\\Clippinger Labs, Ohio University, Athens, OH 45701, USA}

\begin{abstract}

We extend our approach of modeling spectral energy distribution (SED)
and lightcurves of blazars to include external Compton (EC) emission
due to inverse Compton scattering of an external anisotropic target
radiation field. We describe the time-dependent impact of such seed
photon fields on the evolution of multifrequency emission and spectral
variability of blazars using a multi-zone time-dependent leptonic jet
model, with radiation feedback, in the internal shock model scenario.

We calculate accurate EC-scattered high-energy spectra produced by
relativistic electrons throughout the Thomson and Klein-Nishina
regimes. We explore the effects of varying the contribution of (1) a
thermal Shakura-Sunyaev accretion disk, (2) a spherically symmetric
shell of broad-line clouds, the broad line region (BLR), and (3) a hot
infrared emitting dusty torus (DT), on the resultant seed photon
fields. We let the system evolve to beyond the BLR and within the DT
and study the manifestation of the varying target photon fields on the
simulated SED and lightcurves of a typical blazar. The calculations of
broadband spectra include effects of $\gamma-\gamma$ absorption as
$\gamma$-rays propagate through the photon pool present inside the jet
due to synchrotron and inverse Compton processes, but neglect
$\gamma-\gamma$ absorption by the BLR and DT photon fields outside the
jet. Thus, our account of $\gamma-\gamma$ absorption is a lower limit
to this effect. Here, we focus on studying the impact of parameters
relevant for EC processes on high-energy (HE) emission of blazars. 
\end{abstract}


\section{\label{intro}Introduction}

Blazars are known for their highly variable broadband emission. They
are characterized by a doubly humped spectral energy distribution
(SED), attributed to non-thermal emission, and spectral
variability. The SED and variability patterns can be used as key
observational features to place constraints on the nature of the
particle population, acceleration of particles, and the environment
around the jet that is responsible for the observed
emission. Conversely, incorporating the nature of the particle
population and the jet environment, as accurately as possible, in
modeling such observational features can enable us to reach a better
agreement between theoretical and observational results. Thus,
exploring the environment of a blazar jet in an anisotropic and
time-dependent manner is important for connecting the pieces together
and putting tighter constraints on the origin of $\gamma$-ray
emission.

Blazars, a combination of BL Lacertae (BL Lac) objects and flat
spectrum radio-loud quasars (FSRQs), are divided into various
subclasses depending on the location of the peak of the low-energy
(synchrotron) SED component. The synchrotron peak lies in the infrared
regime, with $\nu_{\rm s} \leq 10^{14}$ Hz, in low-synchrotron-peaked
(LSP) blazars comprising FSRQs and low-frequency peaked BL Lac objects
(LBLs). In the case of intermediate-synchrotron-peaked (ISP) blazars,
consisting of LBLs and intermediate-frequency peaked BL Lacs (IBLs),
the synchrotron peak lies in the optical - near-UV region with
$10^{14} < \nu_{\rm s} \leq 10^{15}$ Hz. The synchrotron component of
high-synchrotron-peaked (HSP) blazars, which include essentially all
high-frequency-peaked BL Lac objects (HBLs), peaks in the X-rays at
$\nu_{\rm s} > 10^{15}$ Hz \citep[]{ab2010, bm2012}. The high-energy
(HE) component of blazars can be a result of inverse Compton (IC)
scattering of synchrotron photons internal to the jet resulting in
synchrotron self-Compton (SSC) emission \citep{bm1996}. It could also
be due to upscattering of accretion-disk photons \citep[]{ds1993},
and/or photons initially from the accretion disk being scattered by
the broad-line region (BLR) \citep[]{sbr1994, dss1997}, and/or seed
photons from a surrounding dusty torus (DT) \citep[]{ka1999,
  bl2000}. In the case of HBLs, the HE component is usually well
reproduced with a synchrotron/SSC leptonic jet model
\citep[e.g.,][]{fdb2008, aj2012}, whereas an additional external
Compton (EC) component is almost always required to fit the
high-energy spectra of FSRQs, LBLs, and IBLs \citep[e.g.,][]{cg1999,
  co2010}.

Detailed numerical calculations for Compton scattering processes have
been carried out for many specific models of blazar jet emission that
involve their environment. \citet[]{ds1993, ds2002} have calculated
Compton scattering of target photons in the Thomson regime from an
optically thick and geometrically thin, thermal accretion disk based
on the model of \cite{ss1973}. Quasi-isotropic seed photon fields due
to BLR or DT have also been considered to obtain Compton-scattered
high-energy spectra in the Thomson limit by several authors
\citep[]{sbr1994, dss1997, bl2000}. On the other hand, extensive
calculations involving anisotropic accretion-disk and BLR seed photon
fields have been considered as well \citep[]{bms1997, bb2000, br2004,
  kt2005}. Anisotropic radiation fields of the disk, the BLR, and the
DT have been studied previously by \cite{dp2003}, but primarily in the
context of $\gamma-\gamma$ interaction of these photons with the GeV
and TeV photons produced in the jet. Anisotropic treatment of BLR and
DT photons, focussing on jet emission and rapid non-thermal flares,
was carried out by \cite{sm2005}. These authors studied parameters
describing the properties of BLR and DT that govern the interplay
between the dominance of SSC and EC emission and their subsequent
impact on SEDs, as well as relative time delays between light curves
at different frequencies. For the purposes of their study, they used
an integrated intensity - instead of considering line and continuum
intensities separately - of the incident emission from the BLR. The
emitting plasma was assumed to be located at parsec scales and the
evolution of HE emission at sub-pc distances was ignored.

Recently, anisotropic treatment of disk and BLR target radiation
fields has been considered by \cite{de2009}. The authors have
calculated accurate $\gamma$-ray spectra due to inverse Comptonization
of such seed photon fields throughout the Thomson and Klein-Nishina
(KN) regimes to model FSRQ blazars, although in a one-zone
scenario. Also, these authors evolve the system to only sub-pc
distances along the jet axis, limiting themselves to locations within
the BLR. In addition to this, one-zone leptonic jet models were
recently shown \citep{brm2009} to have severe limitations in attempts
to reproduce very high energy (VHE) flares, such as that of 3C~279
detected in 2006 \citep{al2008}.

In a more recent approach, \cite{ma2013} has considered an anisotropic
seed photon field of the DT to calculate the resultant EC component of
HE emission from blazar jets, in a turbulent extreme multi-zone
scenario. While the $\gamma$-ray spectra are calculated throughout the
Thomson and KN regimes, the energy loss rates are limited to only the
Thomson regime. For the problem that work addresses, the system is
located \textit{beyond} the BLR, at parsec-scale distances from the
central engine.

Here, we extend the previous approach of \citet[][hereafter Paper
  1]{jb2011}, which calculated the synchrotron and SSC emission from
blazar jets, to address some of the limitations of the models
mentioned above. We use a fully time-dependent, 1-D multi-zone with
radiation feedback, leptonic jet model in the internal shock scenario,
shortened to the MUlti ZOne Radiation Feedback, MUZORF, model. We
evolve the system from sub-pc to pc scale distances along the jet
axis. We consider anisotropic target radiation fields to calculate the
HE spectra resulting from EC scattering processes. The entire spectrum
is calculated throughout the Thomson and KN regimes, thereby making it
applicable to all classes of blazars. We include the attenuation of
jet $\gamma$-rays through $\gamma-\gamma$ absorption (described in
Paper 1) due to the presence of target radiation fields inside the
jet, in a self-consistent manner. The generalized approach of our
model lets us account for the constantly changing contribution of each
of the seed photon field sources in producing HE emission in a
self-consistent and time-dependent manner. This is especially relevant
for understanding the origin of $\gamma$-ray emission from blazar
jets.

In a number of previous analyses, the region within the BLR has been
considered the most favorable location for $\gamma$-ray emission, with
a range limited to between 0.01 and 0.3 pc \cite[]{ds1994, bl1995,
  gm1996}. The reason behind this is the short intra-day variability
timescales observed in some $\gamma$-ray flares, which indicated on
the basis of light crossing timescales that the emission region is
small and hence not be too far away from the central engine
\cite[]{gm1996, gt2009}. At the same time, the emission region cannot
be too close to the central engine without violating constraints
placed by the $\gamma-\gamma$ absorption process \cite[]{gm1996,
  lb2006}. As a result, an emission region location closer to the BLR
was considered the most favorable position due to the strong
dependence of the scattered flux on the level of boosting and the
energy of incoming photons \citep{sbr1994}. Contrary to the above
scenario, recent observations have shown coincidences of $\gamma$-ray
outbursts with radio events on pc scales \cite[e.g.,][]{lt2012,
  jo2013}. This seems to suggest a cospatial origin of radio and
$\gamma$-ray events located at such distances. As a result, some
authors conclude that the $\gamma$-ray emitting region could also lie
outside of the BLR \cite[]{sm2005, lvt2005}.

Thus, in order to understand the origin of $\gamma$-ray emission, it
is important to let the system being modeled evolve to beyond the BLR
and into the DT, and to include its contribution to the production of
$\gamma$-ray emission. Here, we focus our attention toward
understanding the dependence of $\gamma$-ray emission on the
combination of various intrinsic physical parameters. We explore this
aspect by including various components of seed photon fields in order
to obtain a complete picture of their contribution in producing
$\gamma$-ray emission and understand their effects on the dynamic
evolution of SEDs and spectral variability patterns.

In \S\ref{method}, we describe our EC framework of including
anisotropic seed photon fields from the accretion-disk, the BLR, and
the DT. We lay out the expressions used to calculate accurate
Compton-scattered $\gamma$-ray spectra resulting from the seed photon
fields and the corresponding electron energy loss rates throughout the
Thomson and KN regimes. In \S\ref{paramstud}, we describe our baseline
model, its simulated results, and the relevant physical input
parameters that we use in the study. In \S\ref{outcome}, we present
our results of the parameter study and discuss the effects of varying
the input parameters on the simulated SED and lightcurves. We discuss
and summarize our findings in \S\ref{disco}. Throughout this paper, we
refer to $\alpha$ as the energy spectral index such that flux density,
$F_{\nu}, \propto \nu^{- \alpha}$; the unprimed quantities refer to
the rest frame of the AGN (lab frame), primed quantities to the
comoving frame of the emitting plasma, and starred quantities to the
observer's frame; the dimensionless photon energy is denoted by
$\epsilon = \frac{h\nu}{m_{\rm e}c^{2}}$.

Appendix \ref{loseqns} delineates the details of line-of-sight
calculations for the BLR line and diffuse continuum emission used in
obtaining the intensity of incoming BLR photons.

\section{\label{method}Methodology}

We consider a multi-zone time-dependent leptonic jet model with
radiation-feedback scheme as described in Paper 1. We extend our
previous model of synchrotron/SSC emission to include the EC component
in order to simulate the SED and spectral variability patterns of
blazars. We consider three sources of external seed photon fields,
namely the accretion disk, the BLR, and the DT. We evolve the emission
region in the jet from sub-pc to pc scales (within the DT) and follow
the evolution of the SED and spectral variability patterns over a
period of $\sim$ 1 day, corresponding to the timescale of a rapid
nonthermal flare. Such a comprehensive approach can be used as an
important tool for connecting the origin of $\gamma$-ray emission of a
flare to its multiwavelength properties.

As in paper 1, we consider a cylindrical emission region for our
current study. We assume the emitting volume to be well collimated out
to pc distances, which is a safe assumption to make based on the work
of \cite{jo2005}, and hence do not consider the effects of adiabatic
expansion on the evolution of the strength of the magnetic field or
the electron population in the emission region. The size of the
emission region is assumed to be small in comparison to the sizes of
and distances to the external seed photon field sources. This way, the
external radiation can be safely assumed to be homogeneous throughout
the emitting plasma, although it is still highly anisotropic in the
comoving frame of the plasma. In our current framework, we do not
simulate radio emission as the calculated flux is well below the
actual radio value. This is because we follow the early phase of
$\gamma$-ray production corresponding to a shock position upto $\sim$
1 pc in the lab frame. During this phase, the emission region is
highly optically thick to GHz radio frequencies.

\begin{figure}
\includegraphics[scale=.3]{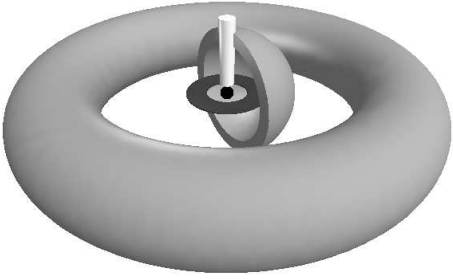}
\caption{Illustration of the three sources of external radiation
  influencing the high-energy emission from the jet of a blazar. The
  central engine and the DT are situated in the same plane while the
  jet and the polar axis of the BLR lie in a plane perpendicular to
  it. Only half of the BLR is shown to illustrate the location of the
  central engine in the spherical shell's cavity.}
\label{jet_geom}
\end{figure}

The angular dependence of the incoming radiation and the amount to
which it contributes toward EC emission is determined by the geometry
of all three seed photon field sources and the location of the
emission region along the jet axis. In addition, the anisotropy is
further enhanced due to relativistic aberration and Doppler boosting
or deboosting in the plasma frame. We assume the external radiation to
be constant in time over the period of our simulation. Figure
\ref{jet_geom} depicts the geometry of all three external sources
under consideration. The jet is oriented along the z-axis in a plane
perpendicular to the plane of the central engine, which is composed of
the black hole and the accretion disk surronding it. The central
engine is enveloped by a BLR, considered to be a geometrically thick
spherical shell, and is situated inside the cavity of the BLR. These
sources are, in turn, encased by a puffed up torus containing hot
dust.

In the following subsections, we discuss the sources of seed photons
for EC scattering and delineate the expressions that we use to
calculate the corresponding emissivities and energy loss rates
throughout the Thomson and KN regimes.

\subsection{\label{disksec}The Accretion Disk}

In order to calculate the EC scattering of photons coming directly
from a central source, we consider an optically thick accretion disk
that radiates with a blackbody spectrum, based on the model of
\cite{ss1973}. The blackbody spectrum is calculated according to a
temperature distribution T(R) given by Eq. (4) of
\cite{bms1997}. where R is the radius of the disk.

\begin{figure}
\includegraphics{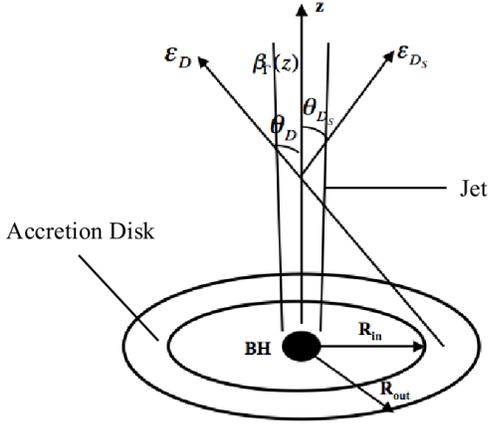}
\caption{Model for the disk-jet geometry used in our EC
  formulation. The accretion disk extends from inner, $R_{\rm in}$, to
  outer, $R_{\rm out}$ radius. Temperature of the disk is calculated
  for all radii between $R_{\rm in}$ and $R_{\rm out}$. The emission
  region is located inside the jet, moving along the z-axis with a
  velocity $\beta_{\Gamma_{\rm sh}} c$ (as described in Paper 1). The
  incoming photon of dimensionless energy $\epsilon_{\rm D}$ from the
  disk intercepts the emitting volume at an angle $\theta_{\rm D}$ and
  the resulting outgoing photon with energy $\epsilon_{\rm D_{\rm S}}$
  is scattered at an angle $\theta_{\rm D_{\rm S}}$.}
\label{diskgeom}
\end{figure}

Figure \ref{diskgeom} shows a schematic of the disk geometry and the
angular dependence of the disk spectral intensity on the position of
the emission region in the jet. We assume a multi-color disk and
calculate the radius dependent quantity, $\Theta(\rm R) \equiv
k_{B}T(R)/m_{e}c^{2}$ (where $k_{B} = 1.38 \times 10^{-16}$~erg/K is
the Boltzmann constant), in order to obtain the EC emissivity and the
corresponding electron energy loss rate. The disk is assumed to emit
in the energy range from optical to hard X-rays (10 keV), with a
characteristic peak frequency of $\nu^{\rm peak}_{\rm disk} \sim 2
\times 10^{15}$~Hz.

 For sake of brevity, the subscript $D$ has been dropped from
 $\epsilon$ for the rest of this section. Now, the spectral surface
 energy flux at radius R is given by $F(\nu, R) (erg~cm^{-2} s^{-1}
 Hz^{-1}) = \pi B_{\nu}[T(R)]$, where $B_{\nu}[T(R)]$ describes the
 spectrum of a blackbody radiation at radius R with temperature T(R)
 \citep{ds1993}. The differential number of photons produced per
 second between $\epsilon$ and $\epsilon + d\epsilon$ and emitted from
 disk radius R and R + dR, $\dot{N}(\epsilon, R)$, is given by

\begin{equation}
\label{diskNeqn}
\frac{dN}{dR dt d\epsilon} = \frac{2 \pi^{2} R B_{\nu}[T(R)]}{h
  \epsilon}~,
\end{equation}
where $B_{\nu}[T(R)] = \frac{2h}{c^{2}}
\frac{\nu^{3}}{\exp{[h\nu/k_{B}T(R)]} - 1}$.

The differential spectral photon number density, $n_{\rm ph}(\epsilon,
R) ~ (cm^{-3} \epsilon^{-1} R^{-1})$, is then given by

\begin{equation}
\label{ndiskeqn}
\frac{dn_{\rm ph}}{d\epsilon dR} = \frac{\dot{N}(\epsilon, R)}{4\pi
  x^{2} c} = \frac{\pi}{2}R\frac{B_{\nu}[T(R)]}{x^{2}ch\epsilon}~,
\end{equation}
where $x = \sqrt{R^{2} + z^{2}}$ and $\cos\theta_{D} = \eta_{\rm ph} =
z/x$. Here and in the rest of the paper, a quantity is differential in
the variables that are listed in parentheses. If the variables are
preceded by a semicolon or only one variable is listed in the
parentheses, then the quantity is parametrically dependent on such
variable(s).
  
Converting $n_{\rm ph}(\epsilon, R)$ into $n_{\rm ph}(\epsilon,
\Omega)$ by realizing that $\eta_{\rm ph} = \frac{z}{\sqrt{R^{2} +
    z^{2}}}$ and assuming azimuthal symmetry of the photon source,
$n_{\rm ph}(\epsilon, \Omega_{\rm ph})$ is given by

\begin{equation}
\label{ndisketaeqn}
\frac{dn_{\rm ph}}{d\epsilon d\Omega_{\rm ph}} =
\frac{1}{2\pi}\frac{\pi B_{\nu}[T(R)]}{2 c h \epsilon \eta_{\rm ph}}~.
\end{equation}

Using Eq. (\ref{ndisketaeqn}) and the invariance of $\frac{n_{\rm
    ph}(\epsilon, \Omega)}{\epsilon^{2}}$ \citep[]{rl1979, ds1993,
  bms1997}, 

\begin{equation}
\label{nepsinvar}
n_{\rm ph}'(\epsilon', \Omega_{\rm ph}') =
\frac{\epsilon'^{2}}{\epsilon^{2}} n_{\rm ph}(\epsilon, \Omega_{\rm
  ph})~,
\end{equation}
we can obtain the anisotropic differential spectral photon number
density, $n_{\rm ph}^{\prime}(\epsilon^{\prime}, \Omega_{\rm
  ph}^{\prime})$, in the plasma frame \citep{bms1997}

\begin{equation}
\label{ndiskcomeqn}
n_{\rm ph}'(\epsilon', \Omega_{\rm ph}') = \frac{1}{2c^{3}}
(\frac{m_{e}c^{2}}{h})^{3} \epsilon'^{2} \left(e^{\frac{\epsilon'
    \Gamma_{\rm sh}(1 + \beta_{\Gamma_{\rm sh}} \eta'_{\rm
      ph})}{\Theta(R)}} - 1\right)^{-1} \frac{1 + \beta_{\Gamma_{\rm
      sh}} \eta'_{\rm ph}}{\eta'_{\rm ph} + \beta_{\Gamma_{\rm sh}}}~,
\end{equation}
Here, $\eta'_{\rm ph} = \frac{z - \beta_{\Gamma_{\rm sh}}x}{x -
  \beta_{\Gamma_{\rm sh}}z}$ is the cosine of the angle that the
incoming photon, emitted at radius $R$, makes with respect to the jet
axis at height z. The relevant Lorentz transformations are given by
\citep{ds1993}

\begin{eqnarray}
\label{lorentzeqn}
\epsilon = \epsilon' \Gamma_{\rm sh} (1 + \beta_{\Gamma_{\rm sh}}
\eta'_{\rm ph}) \nonumber\\ 
\eta_{\rm ph} = \frac{\eta'_{\rm ph} + \beta_{\Gamma_{\rm sh}}}{1 +
  \beta_{\Gamma_{\rm sh}} \eta'_{\rm ph}}~.
\end{eqnarray}

The electron energy loss rate and photon production rate per unit
volume due to inverse-Compton scattering of disk photons (ECD) can be
calculated using Eq. (\ref{ndiskcomeqn}). We use the approximation
given in \cite{bms1997} to calculate the energy loss rate of an
electron with energy $\gamma^{\prime}$ throughout the Thomson and KN
regimes:

\begin{equation}
\label{dgdtecd}
-\dot{\gamma'}_{\rm ECD} = \frac{\pi^{5} r_{e}^{2}}{30c^{2}}
\left(\frac{m_{e}c^{2}}{h}\right)^{3} \gamma^{\prime 2} \Gamma_{\rm
  sh}^{2} \int\limits_{R_{\rm min}}^{R_{\rm max}} dR ~\Theta^{4}(R) ~R
\frac{(x - \beta_{\Gamma_{\rm sh}}z)^2}{x^{4}} I(\langle
\epsilon^{\prime} \rangle, \gamma^{\prime}, \eta_{\rm ph}^{\prime})~,
\end{equation}
where, $I(\epsilon^{\prime}, \gamma^{\prime}, \eta_{\rm ph}^{\prime})$
is given by either equation (15) or (16) of \cite{bms1997} according
to the regime it is being calculated in.

\begin{figure}
\plotone{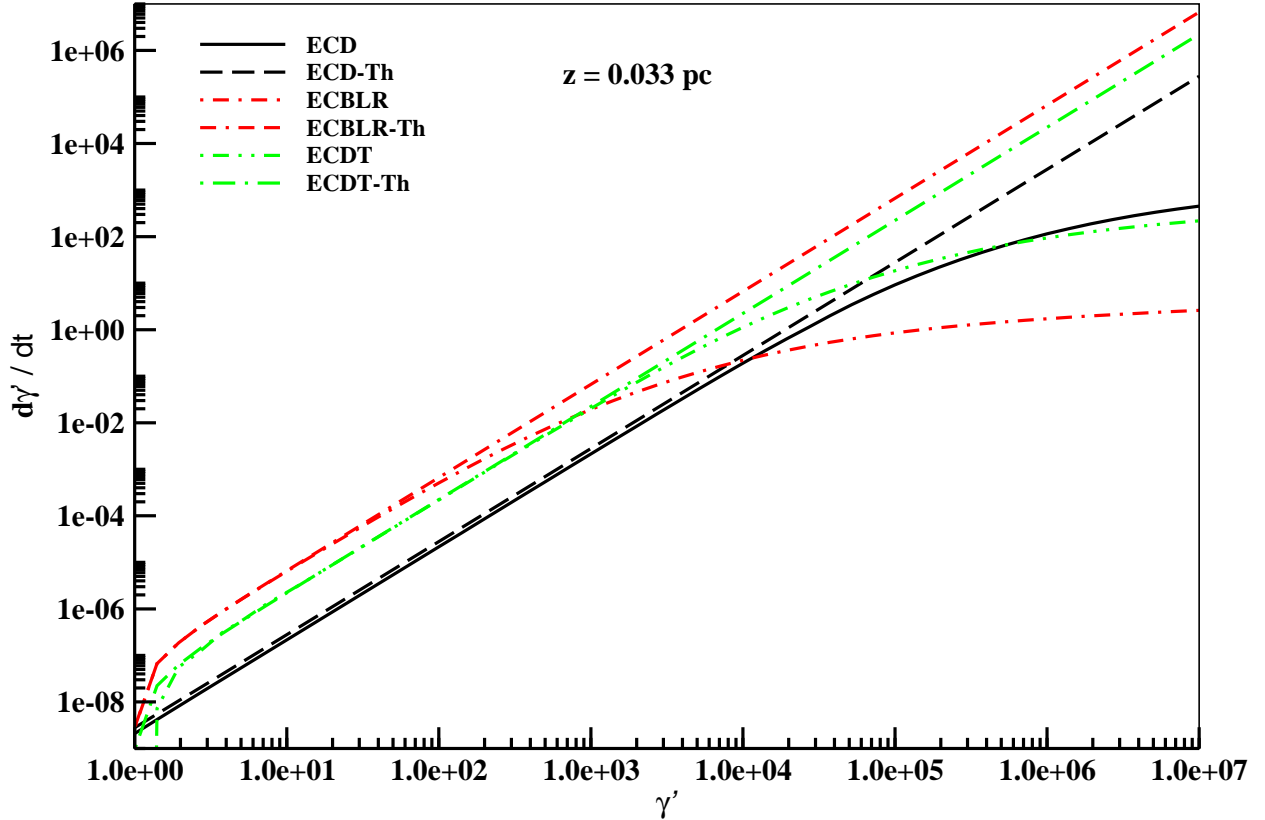}
\caption{Energy loss rates of electrons/positrons showing a comparison
  of Thomson and full expressions due to inverse Comptonization of
  disk photons (ECD), BLR photons (ECBLR), and dusty torus photons
  (ECDT) in the forward shock (FS) region of the emitting volume at a
  distance of 0.033 pc from the BH. Input parameters of the base set
  (see \S\ref{paramstud}) have been used to obtain the respective
  -$\dot{\gamma}^{\prime}$. The ECD process is denoted by solid line
  while its Thomson regime counterpart (ECD-Th) is depicted by
  long-dashed line. The ECBLR process is denoted by dot-dashed line
  and the ECBLR-Th process is represented by dot- double-short-dashed
  line. The ECDT process is represented by double-dot -dashed line and
  the ECDT-Th process is denoted by dot- double-long-dash line.}
\label{dgdtThfull}
\end{figure}

On the other hand, if all scattering occurs in the Thomson regime,
$\gamma' \epsilon' \ll 1$, then the electron energy loss rate can be
directly calculated using Eq. (12) of \cite{bms1997}. Figure
\ref{dgdtThfull} shows a comparison between the electron energy loss
rate obtained using the full expression given in Eq. (\ref{dgdtecd})
and the Thomson expression using Eq. (12) of \cite{bms1997} for
ECD. The transition from the KN to Thomson regime is governed by the
temperature of the accretion disk such that the transition electron
Lorentz factor is given by $\gamma_{\rm KN} \sim 1/(5 \times 10^{-10}
[T_{\rm max}/K])$, where $T_{\rm max} (K) = 0.127
\left[3GM\dot{M}/(8\pi \sigma_{\rm SB} R_{g}^{3})\right]^{1/4}$ is the
maximum temperature of the disk for a non-rotating black hole
(BH). The quantity $R_{\rm g} = \frac{GM}{c^{2}}$ is the gravitational
radius corresponding to a BH of mass M (in units of $M_{\odot}$). The
accretion rate of the BH is given by $\dot{M}$ such that the total
disk luminosity, $L_{\rm disk}$, and the accretion efficiency, $\eta$,
are related to $\dot{M}$ as $L_{\rm disk} = \eta \dot{M} c^{2}$. The
Stefan-Boltzmann constant, $\sigma_{\rm SB} = 5.6704 \times 10^{-5}~
erg ~cm^{-2} K^{-4} s^{-1}$. In the case of our baseline model,
$T_{\rm max} = 9.6 \times 10^{4} K$, implying $\gamma_{\rm KN} \sim 2
\times 10^{4}$. 

As can be seen from the figure, the lines for ECD and ECD-Th do not
overlap each other in the Thomson regime. This is due to the fact that
Eq. (13) of \cite{bms1997} was used to calculate the electron energy
loss rate due to external Comptonization of disk photons. The
expression employs an approximation for all electron energies in
calculating the electron energy loss rate. According to the
approximation, the exact value of the angle between the electron and
the jet axis does not play an important role and could be taken to be
perpendicular to the jet axis for all electron energies.  Furthermore,
the thermal spectrum emitted by each radius of the disk could be
approximated by a delta function in energy. Hence, the resulting
electron energy loss rate is slightly different from that obtained
using the full KN cross-section for an extended source.

The Compton photon production rate per unit volume, in the head-on
approximation with $\beta_{\gamma^{\prime}} \rightarrow 1$, can be
calculated from Eqs. (23) and (25) of \cite{de2009}. Using the
following relationship between spectral luminosity, emissivity, and
photon production rate per unit volume \citep{dm2009}, we obtain

\begin{eqnarray}
\label{luminrel}
\epsilon L(\epsilon, \Omega) = V_{b} \epsilon j(\epsilon,
\Omega) \nonumber\\
\epsilon j(\epsilon, \Omega) = m_{e}c^{2} \epsilon^{2} \dot{n}(\epsilon,
\Omega)~.
\end{eqnarray} 
After substituting the expression from Eq. (\ref{ndiskcomeqn}), and
converting $\eta_{\rm ph}'$ in terms of R as mentioned above, we can
obtain the ECD photon production rate per unit volume, in the plasma
frame, under the head-on approximation as

\begin{eqnarray}
\label{ndotecd}
\dot{n}_{\rm ECD}' (\epsilon_{s}', \Omega_{s}') =
\frac{3\sigma_{T}}{64\pi c^{2}} \frac{1}{\Gamma_{\rm sh}^{2}}
\int\limits_{0}^{2\pi}d\phi'_{\rm ph} \int\limits_{R_{\rm in}}^{R_{\rm
    out}} dR \frac{R}{(x - \beta_{\Gamma_{\rm sh}} z)^{2}} \nonumber\\
\int\limits_{0}^{\epsilon'_{\rm max}} d\epsilon'
\frac{\epsilon'}{e^{\frac{\epsilon' x}{\Theta(R) \Gamma_{\rm sh} (x
      - \beta_{\Gamma_{\rm sh}} z)}} - 1} \int\limits_{\gamma'_{\rm
    low}}^{\infty}d\gamma' \frac{n'_{e}(\gamma')}{\gamma^{'2}}
\Xi_{\rm C}~,
\end{eqnarray}
where the subscript $S$ stands for Compton scattered quantities,
$\sigma_{T} = 6.65 \times 10^{-25} cm^{2}$ is the Thomson
cross-section for an electron, and $n^{\prime}_{e}(\gamma^{\prime})$
is the electron number density. The quantity $\Xi_{C}$ is the
solid-angle integrated KN Compton cross-section, under the head-on
approximation \citep[]{de2009, dm2009} given by

\begin{equation}
\label{xieqn}
\Xi_{c} = \frac{\gamma' - \epsilon_{s}'}{\gamma'} +
\frac{\gamma'}{\gamma' - \epsilon_{s}'} - \frac{2
  \epsilon_{s}'}{\gamma' \epsilon' (1 - \cos\Psi^{\prime}) (\gamma' -
  \epsilon_{s}')} + \frac{\epsilon_{s}^{'2}}{\gamma^{'2} \epsilon^{'2}
  (1 - \cos\Psi')^{2} (\gamma' - \epsilon_{s}')^{2}}
\end{equation}

The quantities $\epsilon^{\prime}_{\rm max}$ and $\gamma^{\prime}_{\rm
  low}$ are given by

\begin{eqnarray}
\label{epsgameqn}
\epsilon'_{\rm max} = \frac{2\epsilon'_{\rm s}}{(1 - \cos\Psi')}
 & \textrm{and} \nonumber\\ 
\gamma'_{\rm low} = \frac{\epsilon'_{\rm S}}{2} \left[1 + \sqrt{1 +
    \frac{2}{\epsilon' \epsilon_{\rm S}^{\prime} (1 - \cos
      \Psi^{\prime})}}\right]~,
\end{eqnarray}
where $\cos \Psi'$, given by Eq. (6) of \cite{bms1997}, is the cosine
of the scattering angle between the electron and target photon
directions. We take $\phi^{\prime}_{\rm e} = 0$ without loss of
generality, based on the assumed azimuthal symmetry of electrons in
the emission region. Eqs. (\ref{ndotecd}), (\ref{ndotblr}) (see
\S\ref{blrsec}), and (\ref{ndotdt}) (see \S\ref{dtsec}) are evaluated
such that in the case of those scatterings for which $\gamma^{\prime}
\epsilon^{\prime} < 0.1$, we use Eq. (44) of \cite{de2009} to
calculate the Compton cross-section in the Thomson regime under
the head-on approximation.

For cases where all scattering occurs in the Thomson regime, we can
substitute the following differential cross-section, in the head-on
approximation, \citep{dm2009}:

\begin{equation}
\label{thomsoneqn}
\frac{d^{2}\sigma_{\rm C}}{d\epsilon^{\prime}_{\rm S}
  d\Omega^{\prime}_{\rm S}} = \sigma_{\rm T}
\delta (\Omega^{\prime}_{\rm S} - \Omega^{\prime}_{\rm e})
\delta \left(\epsilon^{\prime}_{\rm S} - \gamma^{\prime}
\epsilon^{\prime} (1 - \beta_{\gamma^{\prime}}\cos \psi^{\prime})\right)~,
\end{equation}
where the subscript $e$ corresponds to electron related quantities,
and Eq. (\ref{ndiskcomeqn}) in the expression

\begin{equation}
\label{ndoteceqn}
\dot{n}^{\prime}_{\rm EC} (\epsilon^{\prime}_{\rm S},
\Omega^{\prime}_{\rm S}) = \frac{c}{4\pi}
\int\limits_{4\pi}d\Omega^{\prime}_{\rm ph}
\int\limits_{1}^{\infty}d\gamma^{\prime}
n^{\prime}_{e}(\gamma^{\prime})
\int\limits_{0}^{\infty}d\epsilon^{\prime} n^{\prime}_{\rm
  ph}(\epsilon^{\prime}, \Omega^{\prime}_{\rm ph}) (1 -
\beta_{\gamma^{\prime}} \cos \psi^{\prime})
\int\limits_{4\pi}d\Omega^{\prime}_{\rm e} \frac{d^{2}\sigma_{\rm
    C}}{d\epsilon^{\prime}_{\rm S} d\Omega^{\prime}_{\rm S}}~
\end{equation}
to obtain the Thomson regime photon production rate per unit volume
in the plasma frame:

\begin{eqnarray}
\label{ndotecdth}
\dot{n}^{\prime \rm Th} (\epsilon^{\prime}_{\rm S},
\Omega^{\prime}_{\rm S}) = \frac{\sigma_{\rm T}}{8\pi c^{2}}
\left(\frac{m_{e}c^{2}}{h}\right)^{3} \frac{\epsilon^{\prime 2}_{\rm
    S}}{\Gamma^{2}_{\rm sh}} \int\limits_{1}^{\infty}d\gamma^{\prime}
\frac{n^{\prime}_{e}(\gamma^{\prime})}{\gamma^{\prime 6}}
\int\limits_{0}^{2\pi} d\phi^{\prime}_{\rm ph} \frac{1}{1 -
  \beta_{\gamma^{\prime}} \cos \psi^{\prime}}
  \nonumber\\ \int\limits_{R_{\rm in}}^{R_{\rm out}} dR \frac{R}{\left(x -
    \beta_{\Gamma_{\rm sh}} z\right)^{2}}
  \left(e^{\frac{\epsilon^{\prime}_{\rm S} x}{\Gamma_{\rm sh}
      \gamma^{\prime 2} \Theta(R) (x - \beta_{\Gamma_{\rm sh}} z) (1 -
      \beta_{\gamma^{\prime}} \cos \psi^{\prime})}} - 1\right)^{-1}~.
\end{eqnarray}

\subsection{\label{blrsec}The Broad Line Region}

Here we model the BLR as an optically thin and geometrically thick
spherical shell, extending from radius $R_{\rm in, BLR}$ to $R_{\rm
  out, BLR}$, with an optical depth $\tau_{\rm BLR}$ \citep{dp2003}
and a covering factor $f_{\rm cov}$ of the central UV radiation
\citep{lb2006}. We assume the BLR to consist of dense clouds, which
reprocess a fraction of the central UV radiation to produce the broad
emission lines \citep[]{lb2006, de2009}. For our purposes, we assume
that the radial dependence of line emissivity is based on the best fit
parameters (s = 1 and p = 1.5) of \cite{kn1999}, such that the number
density of clouds $n_{c}(r) \propto r^{-1.5}$ and the radius of clouds
$R_{c}(r) \propto r^{1/3}$, at distance $r$ from the BH. In addition,
the BLR clouds Thomson scatter a portion of the central UV radiation
into a diffuse continuum \citep{lb2006}. The line emission and diffuse
continuum can provide important sources of target photons that jet
electrons scatter to produce $\gamma$-ray energies \citep[]{si1997,
  de2009}.

\begin{figure}
\includegraphics[scale=.6]{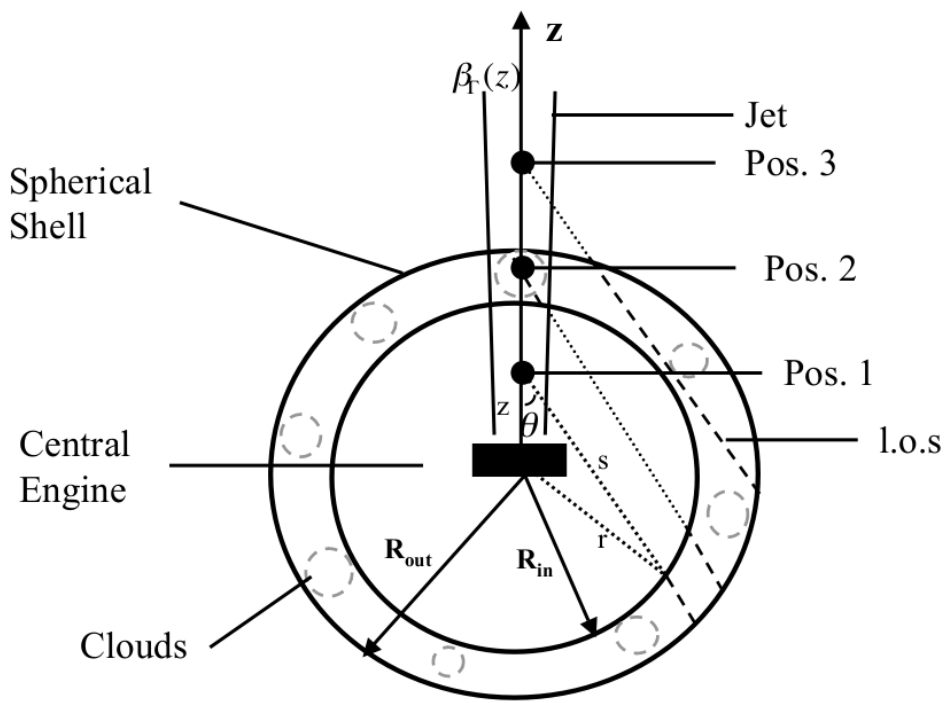}
\caption{Model for the BLR geometry used in our EC component
  calculations. The BLR is considered to be a geometrically thick
  spherical shell that extends from inner, $R_{\rm in, BLR}$, to
  outer, $R_{\rm out, BLR}$, radius. Three positions of the emission
  region are marked: (Pos. 1) region is located in the cavity of the
  BLR, (Pos. 2) region is located within the BLR, and (Pos. 3) region
  is located outside the BLR. The incoming photon makes an angle
  $\theta$ with the z-axis (jet axis). Different lines of sight
  (l.o.s) are shown, which are calculated using the law of cosines
  between $r, s$ and $\theta$ as described in Appendix \ref{loseqns}.}
\label{blrgeom}
\end{figure}

In order to obtain the EC component due to the seed photon field of
the BLR, we need to calculate the anisotropic distribution of the BLR
line and continuum emission. This can be achieved by integrating the
line and continuum emissivities along the lines of sight through the
BLR to obtain the corresponding intensities \citep[]{dp2003,
  lb2006}. We use Eqs. (12) and (13) of \cite{lb2006} to calculate the
anisotropic intensity of radiation of line emission, $I_{\rm line}(z,
\theta)$, and diffuse continuum, $I_{\rm cont}(z, \theta)$ (in units
of $erg ~s^{-1} cm^{-2} ster^{-1}$) as a function of distance $z$ from
the central source and angle $\theta$ that the incoming photons make
with the jet axis. Figure \ref{blrgeom} represents the geometry of the
BLR under consideration and the angular dependence of the intensity of
radiation from the BLR at the position of the emission region in the
jet. We consider three possible positions of the emission region
\citep{dp2003} to calculate emissivities and corresponding
intensities, $I(z, \mu)$, along the jet axis, where $\mu$ is the
cosine of the angle $\theta$ that the incoming BLR photon makes with
the jet axis. The calculations of these path lengths are described in
Appendix \ref{loseqns}.

\begin{figure}
\plottwo{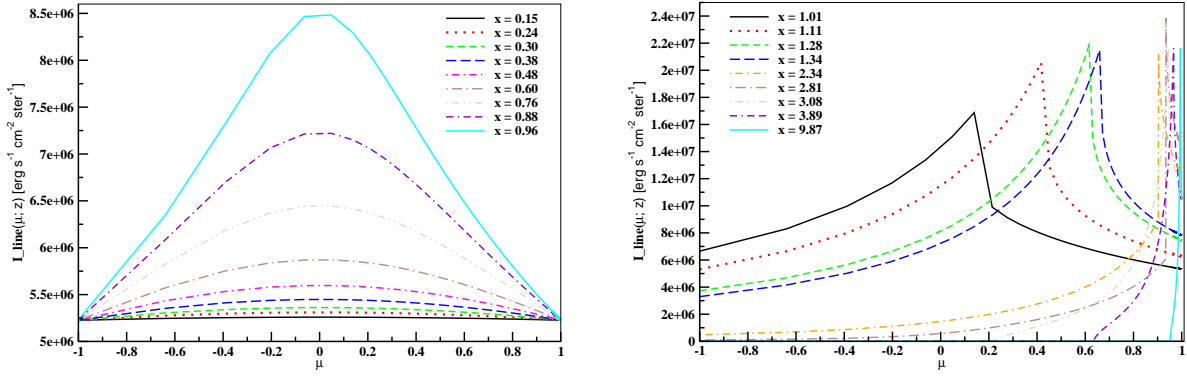}{f7}
\caption{Plot of intensity of radiation as a function of $\mu =
  \cos{\theta}$, of BLR line emission. The variable $x$ in the figures
  refers to the distance along the jet axis in units of $R_{\rm in,
    BLR}$. Input parameters of the base set (see \S\ref{paramstud})
  are used to obtain the intensity profile of the BLR emission
  lines. Left: Intensity profile when the emission region is located
  at Pos. 1. The profile is symmetric due to equal contribution of all
  l.o.s from the BLR. The profile peaks as the emission region
  approaches the inner radius of the BLR at $\sim 0.96 R_{\rm in,
    BLR}$. Right: Intensity profile when the emission region is
  located at Pos. 2 and 3. The profile becomes asymmetric as the
  emission region moves to Pos. 2 and Pos. 3 because lines of sight of
  unequal lengths contribute to the intensity calculation. The
  intensity distribution peaks at Pos. 2 when the emission region is
  located within the BLR shell, at $\sim 2.81 R_{\rm in, BLR}$ . The
  intensity plummets as the emission region emerges from the BLR
  shell, at $\sim 3.08 R_{\rm in, BLR}$, and stays constant
  thereafter.}
\label{Izprofile}
\end{figure}

The anisotropic profile of emission line intensity obtained using the
path length calculations, as described in Appendix \ref{loseqns}, at
the three locations (marked in Figure \ref{blrgeom}) is shown in
Figure \ref{Izprofile}. The anisotropic intensity due to diffuse BLR
emission has a similar profile as that of emission line intensity, and
is not shown here for the sake of brevity.

For the purposes of our study, we consider both the broad line
emission and the diffuse continuum radiation to calculate the total
radiation field of the BLR. The combined field provides the source of
target photons for EC scattering (ECBLR) by jet electrons. The BLR is
assumed to emit in the energy range from infrared (IR) to soft X-rays
(3 keV), with a characteristic peak frequency of $\nu^{\rm peak}_{\rm
  BLR} \sim 2 \times 10^{15}$~ Hz. For the sake of brevity, we drop
the subscript $BLR$ from the equations for the rest of this section.

We consider 35 emission lines (34 components from \cite{fr1991} and
the H$\alpha$ component from \cite{gsw1981}) to estimate the total
flux of broad emission lines. Using Eq. (19) of \cite{lb2006} and
Eq. (\ref{nepsinvar}), we can obtain the differential line emission
photon number density (in $cm^{-3} \epsilon^{-1} ster^{-1}$) in the
plasma frame:

\begin{equation}
\label{nlinecomeqn}
n^{\prime \rm line}_{\rm ph} (\epsilon^{\prime}_{\rm line},
\Omega^{\prime}_{\rm ph}; z) = C_{1} \frac{I_{\rm line}(z,
  \mu)}{\Gamma_{\rm sh}^{4}(1 + \beta_{\Gamma_{\rm sh}}
  \mu^{\prime})^{4}} \frac{N_{\rm V} ~\delta(\epsilon^{\prime} -
  \epsilon^{\prime}_{\rm line})}{\epsilon^{\prime}}~,
\end{equation}
where $C_{1} = \frac{1}{m_{e} c^{3} 555.77}$, $\epsilon^{\prime}_{\rm
  line}$ is the dimensionless energy of the incoming photon
corresponding to one of the 35 emission line components and $N_{\rm
  V}$ is the line strength of each of those 35 components, with that
of Ly$\alpha$ arbitrarily set at 100 \citep{fr1991}. We define
$\mu^{\prime}$ from $-1$ to $1$ in the plasma frame and use
Eq. (\ref{lorentzeqn}) to obtain $\mu$ for the lab frame.

Similarly, using Eqs. (19), (20), and (21) of \cite{lb2006} and
Eq. (\ref{nepsinvar}), we can obtain the differential diffuse
continuum photon number density (in $cm^{-3} \epsilon^{-1} ster^{-1}$)
in the plasma frame as

\begin{equation}
\label{ncontcomeqn}
n^{\prime \rm cont}_{\rm ph} (\epsilon^{\prime}_{\rm cont},
\Omega^{\prime}_{\rm ph}; z) = C_{2} \frac{I_{\rm cont}(z, \mu)
  \epsilon^{\prime 2}}{\left(e^{\frac{\epsilon^{\prime} \Gamma_{\rm sh} (1
      + \beta_{\Gamma_{\rm sh}} \mu^{\prime})}{\Theta}} - 1\right) I}~,
\end{equation}
where $C_{2} = \frac{1}{m_{e}c^{3}}$ and $\Theta = 1.68 \times
10^{-5}$, corresponding to a blackbody temperature of $T = 10^{5}$ K,
which has been assumed for the inner region of the accretion disk
\citep{lb2006}. The quantity $I$ is the total blackbody spectrum,
given by

\begin{equation}
\label{contlabeqn}
I = \int\limits_{\epsilon_{\rm min}}^{\epsilon_{\rm max}}
\frac{\epsilon^{3} d\epsilon}{e^{\frac{\epsilon}{\Theta}} - 1}~,
\end{equation}
with $\epsilon_{\rm min} = 3.22 \times 10^{-6}$ corresponding to the
photon frequency $\nu_{\rm min} = 10^{14.6}$ Hz, and $\epsilon_{\rm
  max} = 2.56 \times 10^{-4}$ corresponding to the frequency $\nu_{\rm
  max} = 10^{16.5}$ Hz \citep{lb2006}. Thus, the total anisotropic
differential spectral photon number density entering the jet from the
BLR is

\begin{equation}
\label{nblrcomeqn}
n_{\rm ph}^{\prime} (\epsilon^{\prime}, \Omega^{\prime}_{\rm ph}) =
n^{\prime \rm line}_{\rm ph} (\epsilon^{\prime}, \Omega^{\prime}_{\rm
  ph}) + n^{\prime \rm cont}_{\rm ph} (\epsilon^{\prime},
\Omega^{\prime}_{\rm ph})~.
\end{equation}

The electron energy loss rate and photon production rate per unit
volume due to ECBLR can be calculated using Eqs. (\ref{nlinecomeqn}),
(\ref{ncontcomeqn}) and (\ref{nblrcomeqn}). We use Eq. (6.46) of
\cite{dm2009} to obtain the electron energy loss rate in the plasma
frame. Substituting Eqs. (6.39) and (6.40) in Eq. (6.46) of
\cite{dm2009} yields

\begin{eqnarray}
\label{dgdteceqn}
\dot{\gamma^{\prime}} = \frac{-3c \sigma_{\rm T}}{8} \int\limits_{4\pi}
d\Omega^{\prime}_{\rm ph} \int\limits_{0}^{\infty} d\epsilon^{\prime}
n_{\rm ph}^{\prime} (\epsilon^{\prime}, \Omega^{\prime}_{\rm ph}) 
\{ \frac{\ln(D)}{\gamma^{\prime} \epsilon^{\prime} M^{2}}
  \left[ (\gamma^{\prime} - \epsilon^{\prime}) (M (M - 2) - 2) -
    \gamma^{\prime} \right] + \nonumber\\
  \frac{1}{3 \epsilon^{\prime}D^{3}} [ 1 - D^{3} +
    6 \epsilon^{\prime} D (\gamma^{\prime} - \epsilon^{\prime}) (1 + M)
    (1 - \beta_{\gamma^{\prime}}\mu^{\prime})
    + \nonumber\\ \frac{6 D^{2}}{\gamma^{\prime} M} \left( 2 (\gamma^{\prime} -
    \epsilon^{\prime}) D - \gamma^{\prime} (M (M - 1) -
    1) \right) ] \}~,
\end{eqnarray}
where $M = \gamma^{\prime} \epsilon^{\prime} (1 -
\beta_{\gamma^{\prime}} \mu^{\prime})$ and $D = 1 + 2M$. As can be
seen from the above equation, the entire integral is independent of
$\phi^{\prime}$ and can be solved analytically. Similarly, after
substituting Eqs. (\ref{nlinecomeqn}) and (\ref{ncontcomeqn}) in
Eq. (\ref{dgdteceqn}), the $d\epsilon^{\prime}$ integral can be solved
analytically for $n^{\prime \rm line}_{\rm ph} (\epsilon^{\prime},
\Omega^{\prime}_{\rm ph})$ due to the presence of the $\delta
(\epsilon^{\prime} - \epsilon^{\prime}_{\rm line})$ function in its
expression (see Eq. \ref{nlinecomeqn}). After having carried out these
simplifications, Eq. (\ref{dgdteceqn}) is solved numerically to obtain
the final electron energy loss rate due to the ECBLR process.

In the case that all scattering occurs in the Thomson regime, $\gamma'
\epsilon' \ll 1$, the electron energy loss rate can be directly
calculated by substituting $\langle \epsilon_{S}^{0} \sigma \rangle =
\sigma_{T}$ and $\langle \epsilon_{S}^{1} \sigma \rangle = \sigma_{T}
\gamma^{\prime 2} \epsilon^{\prime} (1 - \cos \psi^{\prime})$ into
Eq. (6.46) of \cite{dm2009}. After carrying out integrations over
$d\phi^{\prime}$ and $d\epsilon^{\prime}$ analytically for both
$n^{\prime \rm line}_{\rm ph} (\epsilon^{\prime}, \Omega^{\prime}_{\rm
  ph})$ and $n^{\prime \rm cont}_{\rm ph} (\epsilon^{\prime},
\Omega^{\prime}_{\rm ph})$, the Thomson-regime electron energy loss
rate expression for the ECBLR process is given by

\begin{equation}
\label{dgdtblrth}
\dot{\gamma^{\prime}} = -2\pi c \sigma_{\rm T} C_{2} \int\limits_{-1}^{1}
d\mu^{\prime} \frac{(1 - \beta_{\gamma^{\prime}} \mu^{\prime})
  (\gamma^{\prime 2} (1 - \beta_{\gamma^{\prime}} \mu^{\prime}) -
  1)}{\Gamma_{\rm sh}^{4} \left(1 + \beta_{\Gamma_{\rm sh}}
  \mu^{\prime}\right)^{4}} \left[I_{\rm line}(z, \mu) + \frac{\left(\pi
      \Theta\right)^{4}}{15 I} I_{\rm cont}(z, \mu)\right]~.
\end{equation}

Here, we have used the result $\int\limits_{0}^{\infty} dx
\frac{x^{3}}{e^{xa} - 1} = \frac{\pi^{4}}{15 a^{4}}$. As can be seen
from Fig. \ref{dgdtThfull}, the Thomson approximation for ECBLR
deviates from the corresponding full expression at $\gamma^{\prime}
\geq 100$.

The ECBLR photon production rate per unit volume, under the head-on
approximation, is calculated using Eq. (6.32) of \cite{dm2009}. We
write it in terms of the differential photon production rate using
Eq. (\ref{luminrel}), and substitute the expression for the
differential photon number density of BLR photons from
Eq. (\ref{nblrcomeqn}) to obtain

\begin{equation}
\label{ndotblr}
\dot{n^{\prime}} (\epsilon_{s}', \Omega_{s}') = \frac{3 c
  \sigma_{T}}{32 \pi} \int\limits_{0}^{2\pi}d\phi'_{\rm ph}
\int\limits_{-1}^{1} d\mu^{\prime} \int\limits_{0}^{\epsilon'_{\rm
    max}} d\epsilon' \frac{n_{\rm ph}^{\prime} (\epsilon^{\prime},
  \Omega^{\prime}_{\rm ph})}{\epsilon^{\prime}}
\nonumber\\ \int\limits_{\gamma'_{\rm low}}^{\infty}d\gamma'
\frac{n'_{e}(\gamma')}{\gamma^{'2}} \Xi_{\rm C}~,
\end{equation}
where the quantities used in the above equation have been explained in
\S\ref{disksec}. As mentioned in \S\ref{disksec}, the Compton
cross-section in the above expression is evaluated such that in the
case of scatterings for which $\gamma^{\prime} \epsilon^{\prime} <
0.1$, we use Eq. (44) of \cite{de2009} to calculate it in the Thomson
regime, under head-on approximation.

For cases where all scattering occurs in the Thomson regime,
Eqs. (\ref{thomsoneqn}) and (\ref{ndoteceqn}) yield

\begin{eqnarray}
\label{ndotecth}
\dot{n}^{\prime \rm Th} (\epsilon^{\prime}_{\rm S},
\Omega^{\prime}_{\rm S}) = \frac{c \sigma_{\rm T}}{4 \pi}
\int\limits_{4 \pi} d \Omega^{\prime}_{\rm ph}
\int\limits_{1}^{\infty} d\gamma^{\prime}
n^{\prime}_{e}(\gamma^{\prime}) (1 - \beta_{\gamma^{\prime}} \cos
\psi^{\prime}) \int\limits_{0}^{\infty} d\epsilon^{\prime}
n^{\prime}_{\rm ph}(\epsilon^{\prime}, \Omega^{\prime}_{\rm ph})
\nonumber\\ \delta \left(\epsilon^{\prime}_{\rm S} - \gamma^{\prime 2}
\epsilon^{\prime} \left[1 - \beta_{\gamma^{\prime}} \cos
  \psi^{\prime}\right]\right)~.
\end{eqnarray}

Solving for $d \epsilon^{\prime}$ analytically, we obtain the Thomson
regime photon production rate per unit volume as

\begin{equation}
\label{ndotectheqn}
\dot{n}^{\prime \rm Th} (\epsilon^{\prime}_{\rm S}, \Omega^{\prime}_{\rm
  S}) = \frac{c \sigma_{\rm T}}{4 \pi} \int\limits_{4 \pi} d
\Omega^{\prime}_{\rm ph} \int\limits_{1}^{\infty} d\gamma^{\prime}
\frac{n^{\prime}_{e}(\gamma^{\prime})}{\gamma^{\prime 2}}
n^{\prime}_{\rm ph}\left(\frac{\epsilon^{\prime}_{\rm
    S}}{\gamma^{\prime 2} (1 - \beta_{\gamma^{\prime}} \cos
  \psi^{\prime})}, \Omega^{\prime}_{\rm ph}\right)~.
\end{equation}

Now substituting Eqs. (\ref{nlinecomeqn}) and (\ref{ncontcomeqn}) in
Eq. (\ref{ndotectheqn}) and solving for the delta function (present in
Eq. (\ref{nlinecomeqn})) in the $d\gamma^{\prime}$ integral for the
line emission part, we can obtain the final expression for the ECBLR
Thomson-regime photon production rate per unit volume as

\begin{eqnarray}
\label{ndotblrtheqn}
\dot{n}^{\prime \rm Th} (\epsilon^{\prime}_{\rm S},
\Omega^{\prime}_{\rm S}) = \frac{c \sigma_{\rm T}}{4 \pi}
\int\limits_{4 \pi} d\Omega^{\prime}_{\rm ph} [ \frac{C_{1}}{2}
  \frac{I_{\rm line}(z, \mu)}{\Gamma_{\rm sh}^{4} \left(1 +
    \beta_{\Gamma_{\rm sh}} \mu^{\prime}\right)^{4}} \sqrt{\frac{1 -
      \cos \psi^{\prime}}{\epsilon^{\prime}_{\rm S}}}
  \sum_{\epsilon^{\prime}_{\rm line}} \frac{N_{\rm
      V}}{\epsilon^{\prime 3/2}_{\rm line}}
  n_{e}\left(\sqrt{\frac{\epsilon^{\prime}_{\rm
        S}}{\epsilon^{\prime}_{\rm line} (1 - \cos
      \psi^{\prime})}}\right) \nonumber\\ + C_{2} \frac{I_{\rm cont}(z,
    \mu) \epsilon^{\prime 2}_{\rm S}}{I} \int\limits_{1}^{\infty}
  d\gamma^{\prime}
  \frac{n^{\prime}_{e}(\gamma^{\prime})}{\gamma^{\prime 6} \left(1 -
    \beta_{\gamma^{\prime}} \cos \psi^{\prime}\right)^{2}}
  \left(e^{\frac{\epsilon^{\prime}_{\rm S} \Gamma_{\rm sh} (1 +
        \beta_{\Gamma_{\rm sh}} \mu^{\prime})}{\gamma^{\prime 2} (1 -
        \beta_{\gamma^{\prime}} \cos \psi^{\prime}) \Theta}} -
    1\right)^{-1} ]~.
\end{eqnarray}
Where we evaluate the line emission term of the above equation for
cases where $\gamma^{\prime} \geq 10$ so that $\beta_{\gamma^{\prime}}
\simeq 1$ and the $d\gamma^{\prime}$ integral can be solved
analytically using the delta function that is present in the
expression for the differential line emission photon number density.

\subsection{\label{dtsec}The Dusty Torus}

\begin{figure}
\includegraphics[scale=.5]{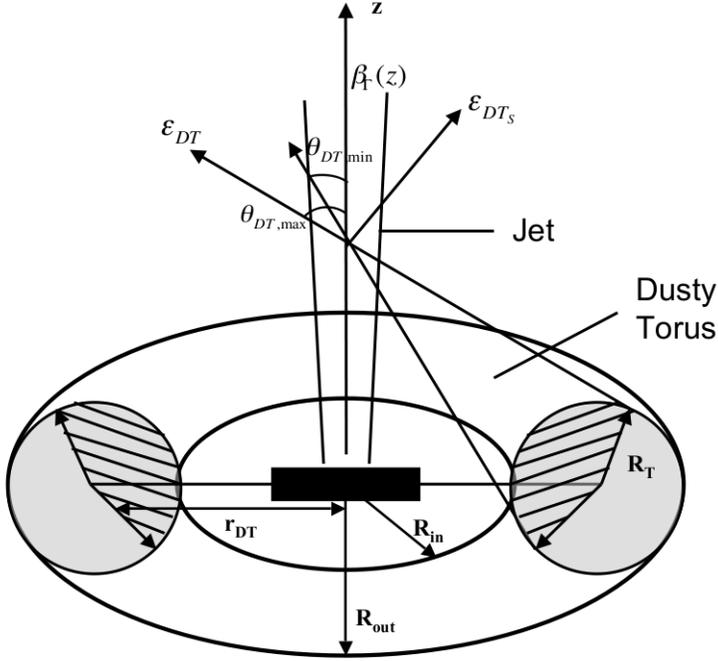}
\caption{Model for the DT geometry used in our EC component
  calculations. The DT is considered to be a patchy molecular torus
  that extends from inner, $R_{\rm in, DT}$, to outer, $R_{\rm out,
    DT}$ radius. The central circle of the torus is located at a
  distance $r_{\rm DT}$ from the central engine. The shaded portion
  shows the cross-sectional area of the torus with radius $R_{\rm
    T}$. Only a fraction of the torus facing the central continuum
  source (cross-hatched region) is assumed to be hot enough to be
  visible to the emission region on the z-axis. Incoming photons from
  this portion of the surface enter the jet at angles whose range is
  governed by a minimum, $\theta_{\rm DT, min}$, and a maximum,
  $\theta_{\rm DT, max}$. The dimensionless energy of an incoming
  photon is given by $\epsilon_{\rm DT}$, while that of the scattered
  or the outgoing photon is given by $\epsilon_{\rm DT, S}$.}
\label{dtgeom}
\end{figure}

We consider a clumpy molecular torus \citep[]{sm2005, ma2013} whose
emission is dominated by dust and which radiates as a blackbody at IR
frequencies at a temperature T = 1200 K \citep{ma2011} in the lab
frame. The torus lies in the plane of the accretion disk and extends
from $R_{\rm in, DT}$ to $R_{\rm out, DT}$. As shown in
Fig. \ref{dtgeom}, the central circle of the torus is located at a
distance of $r_{\rm DT} = \frac{R_{\rm out, DT} + R_{\rm in, DT}}{2}$
from the central source and the cross-sectional radius of the torus is
given by $R_{\rm T} = \frac{R_{\rm out, DT} - R_{\rm in, DT}}{2}$. We
assume that the incident radiation comes from a portion of the inner
surface of the torus. This portion, which is the covering factor, is
dependent on the size of the torus.

The DT is assumed to emit IR photons with a characteristic peak
frequency of the radiation field, $\nu^{\rm peak}_{\rm DT} \sim 3
\times 10^{13}$~ Hz. For the sake of brevity, we drop the subscript
$DT$ from all quantities listed in this section. The minimum,
$\theta_{\rm min}$, and maximun, $\theta_{\rm max}$, incident angles
constraining the incident emission from the torus are given by

\begin{eqnarray}
\label{dtangleseqn}
\theta_{\rm min} = \sin^{-1}{\frac{r}{\sqrt{z^{2} + r^{2}}}} -
\sin^{-1}{\frac{R_{\rm T}}{\sqrt{z^{2} + r^{2}}}}
\nonumber\\ \theta_{\rm max} = \sin^{-1}{\frac{r}{\sqrt{z^{2} + r^{2}}}}
+ \sin^{-1}{\frac{R_{\rm T}}{\sqrt{z^{2} + r^{2}}}}~.
\end{eqnarray}

These angles are subsequently transformed into the plasma frame
according to:

\begin{equation}
\label{comangleeqn} 
\eta^{\prime} = \cos{\theta^{\prime}} = \frac{\cos{\theta} -
  \beta_{\Gamma_{\rm sh}}}{1 - \beta_{\Gamma_{\rm sh}} \cos{\theta}}~.
\end{equation}

The covering factor of the dusty torus, $f_{\rm cov}$, can be obtained
in terms of the fraction, $\xi$, of the disk luminosity, $L_{\rm
  disk}$, that illuminates the torus such that, $L_{\rm Dust} = \xi
L_{\rm disk}$. Here we take $\xi = 0.22$ as found for PKS 1222+216 by
\cite{ma2011}. Also, the following relationship holds between $L_{\rm
  Dust}$, $f_{\rm cov}$, and the illuminated area of the torus,
$A_{\rm obs}$:

\begin{equation}
\label{ldteqn}
L_{\rm Dust} = A_{\rm obs} \sigma_{\rm SB} T^{4} f_{\rm cov}~,
\end{equation}
The illuminated area of the torus visible from a position in the jet
is given by

\begin{equation}
\label{areadteqn}
A_{\rm obs} \approx \frac{\pi^{2}}{4} (R_{\rm out}^{2} - R_{\rm
  in}^{2})~,
\end{equation}
where the factor of 1/4 appears because only the front side of the
inner torus is illuminated and only half of this is visible to the
emitting region in the lab frame. Thus, for given values of $R_{\rm
  in}$, $R_{\rm out}$, and $L_{\rm disk}$, the covering factor of the
torus can be obtained from Eqs. (\ref{ldteqn}) and
(\ref{areadteqn}). Conversely, for given values of $f_{\rm cov}$ and
$R_{\rm in}$, we can also obtain the extent of the torus in terms of
$R_{\rm out}$ and the corresponding values of $r$ and $R_{\rm T}$.

\begin{figure}
\plotone{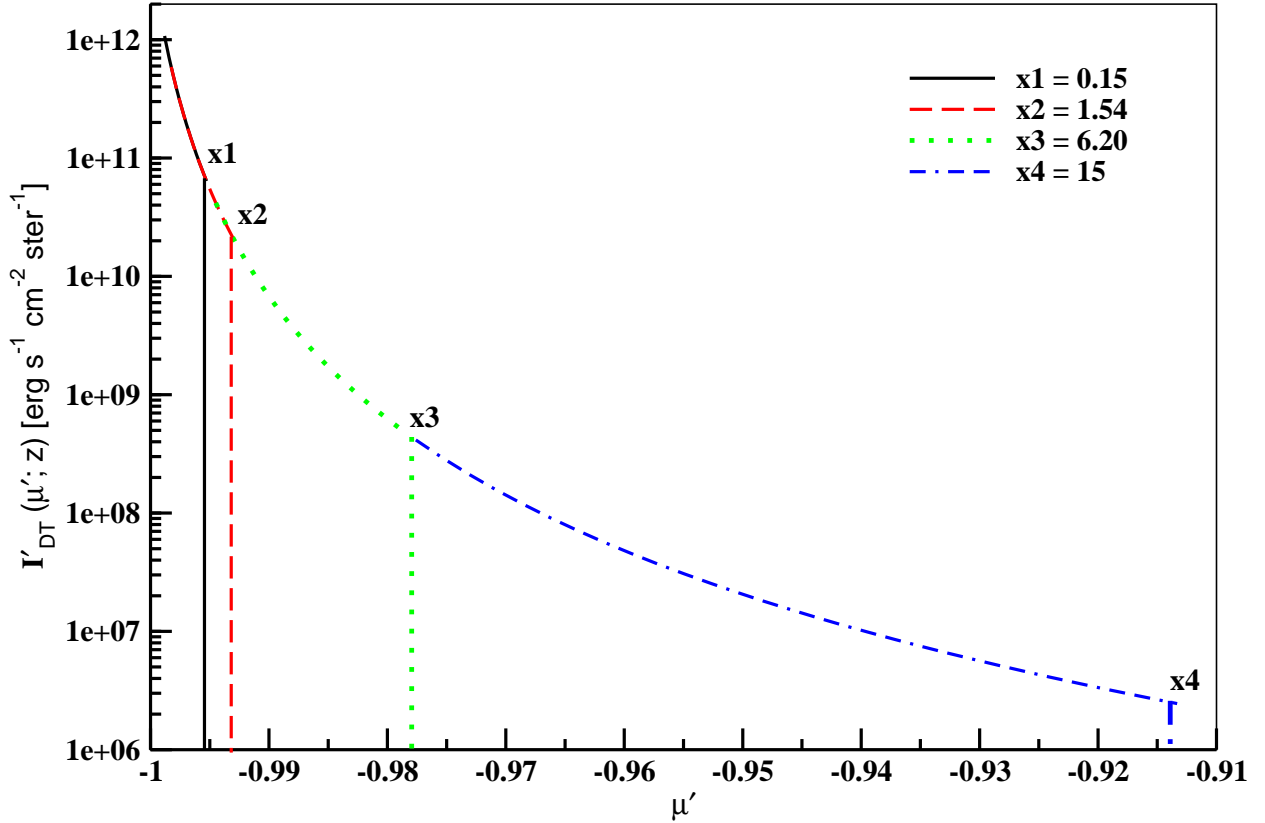}
\caption{Plot of intensity of radiation from the DT as a function of
  distance, $z$, along the jet axis and the incident angle of incoming
  photons $\mu^{\prime} = \cos{\theta^{\prime}}$. The variable $x$ in
  the figure refers to the value of $z$ in terms of $R_{\rm in,
    BLR}$. Input parameters of the base set (see \S\ref{paramstud})
  have been used to obtain the anisotropic intensity profile of the DT
  emission. Since only a fraction of the torus emits radiation in the
  direction of the emission region, at smaller values of $z$, incoming
  photons enter the jet from the front but cover a very narrow range
  of angles. As the emission region moves outward along the jet axis
  (e.g., $z = 6.20 R_{\rm in BLR}$), incoming photons from the DT
  cover a broader range of angles but enter the jet more from the
  side, which leads to de-boosting and reduces the overall intensity
  of radiation in the plasma frame. The dotted lines in the figure
  indicate the range of angles that contribute to the DT intensity at
  that distance; beyond this range the intensity drops to zero.}
\label{dtintenprof}
\end{figure}

Since the torus emits as a blackbody, the differential spectral photon
number density in the plasma frame is simply given by

\begin{equation}
\label{ndtcomeqn}
n^{\prime}_{\rm ph} (\epsilon^{\prime}, \Omega^{\prime}_{\rm ph}) = 2
\left(\frac{m_{e}c}{h}\right)^{3} f_{\rm cov} \frac{\epsilon^{\prime
    2}}{e^{\frac{\epsilon^{\prime} \Gamma_{\rm sh} (1 +
      \beta_{\Gamma_{\rm sh}} \eta^{\prime})}{\Theta}} - 1}~,
\end{equation}
where we have used Eq. (\ref{nepsinvar}) to convert the differential
photon density from the lab to the plasma
frame. Fig. \ref{dtintenprof} shows the anisotropic intensity profile
of the DT as a function of incident angle, $\mu^{\prime} =
\cos{\theta^{\prime}}$, in the plasma frame.

We substitute Eq. (\ref{ndtcomeqn}) in Eq. (\ref{dgdteceqn}) to
calculate the electron energy loss rate due to ECDT, which yields

\begin{eqnarray}
\label{dgdtecdteqn}
\dot{\gamma^{\prime}} = \frac{-3c \pi \sigma_{\rm T}}{2}
\left(\frac{m_{e}c}{h}\right)^{3} f_{\rm cov}
\int\limits_{\eta^{\prime}_{\rm min}}^{\eta^{\prime}_{\rm max}}
d\eta^{\prime} \int\limits_{0}^{\infty} d\epsilon^{\prime}
\frac{\epsilon^{\prime 2}}{e^{\frac{\epsilon^{\prime} \Gamma^{\rm sh}
      (1 + \beta_{\Gamma_{\rm sh}} \eta^{\prime})}{\Theta}} - 1}
\{ \frac{\ln(D)}{\gamma^{\prime} \epsilon^{\prime} M^{2}}
  \left[(\gamma^{\prime} - \epsilon^{\prime})(M (M-2) -
    2)-\gamma^{\prime}\right] \nonumber\\ +
  \frac{1}{3\epsilon^{\prime}D^{3}} [ 1 - D^{3} +
    6\epsilon^{\prime} D (\gamma^{\prime} - \epsilon^{\prime}) (1+M)
    (1-\beta_{\gamma^{\prime}}\eta^{\prime}) + \nonumber\\ \frac{6
      D^{2}}{\gamma^{\prime} M} \left( 2(\gamma^{\prime} -
    \epsilon^{\prime}) D - \gamma^{\prime}(M (M-1) -
    1) \right) ] \}.
\end{eqnarray}

For cases where all scattering occurs in the Thomson regime, we follow
the steps described in \S\ref{blrsec} to obtain Eq. (\ref{dgdtblrth}),
which yields the Thomson-regime electron energy loss rate for the ECDT
process as

\begin{equation}
\label{dgdtdtth}
\dot{\gamma^{\prime}} = \frac{-4 \pi^{5} c \sigma_{\rm T} \Theta^{4}}{15
  \Gamma_{\rm sh}^{4}} \left(\frac{m_{e}c}{h}\right)^{3} f_{\rm cov}
\int\limits_{\eta^{\prime}_{\rm min}}^{\eta^{\prime}_{\rm max}}
d\eta^{\prime} \frac{(1 - \beta_{\gamma^{\prime}} \eta^{\prime})
  \left[\gamma^{\prime 2} (1 - \beta_{\gamma^{\prime}} \eta^{\prime})
    - 1\right]}{\left(1 + \beta_{\Gamma_{\rm sh}}
  \eta^{\prime}\right)^{4}}~.
\end{equation}

As shown in Fig. \ref{dgdtThfull}, the Thomson approximation for ECDT
deviates from the corresponding full expression above $\gamma^{\prime}
\sim 3 \times 10^{3}$. We substitute Eq. (\ref{ndtcomeqn}) in
Eq. (\ref{ndotblr}) to obtain the ECDT photon production rate per unit
volume, under the head-on approximation, as

\begin{equation}
\label{ndotdt}
\dot{n^{\prime}} (\epsilon_{s}', \Omega_{s}') = \frac{3 c \sigma_{\rm
    T}}{16 \pi} \left(\frac{m_{e}c}{h}\right)^{3} f_{\rm cov}
\int\limits_{0}^{2\pi}d\phi'_{\rm ph} \int\limits_{\eta^{\prime}_{\rm
    min}}^{\eta^{\prime}_{\rm max}} d\eta^{\prime}
\int\limits_{0}^{\epsilon^{\prime}_{\rm max}} d\epsilon'
\frac{\epsilon^{\prime}}{e^{\frac{\epsilon^{\prime} \Gamma_{\rm sh}
      (1 + \beta_{\Gamma_{\rm sh}} \eta^{\prime})}{\Theta}} - 1}
\nonumber\\ \int\limits_{\gamma'_{\rm low}}^{\infty}d\gamma'
\frac{n'_{e}(\gamma')}{\gamma^{'2}} \Xi_{\rm C}~.
\end{equation}

For scatterings occuring entirely in the Thomson regime, we substitute
Eq. (\ref{ndtcomeqn}) in Eq. (\ref{ndotecth}) to obtain the ECDT
Thomson-regime photon production rate per unit volume, 

\begin{equation}
\label{ndotdtth}
\dot{n}^{\prime \rm Th} (\epsilon^{\prime}_{\rm S}, \Omega^{\prime}_{\rm
  S}) = \frac{c \sigma_{\rm T}}{2 \pi}
\left(\frac{m_{e}c}{h}\right)^{3} f_{\rm cov} \int\limits_{0}^{2 \pi}
d \phi^{\prime} \int\limits_{\eta^{\prime}_{\rm
    min}}^{\eta^{\prime}_{\rm max}} d\eta^{\prime}
\int\limits_{1}^{\infty} d\gamma^{\prime}
\frac{n^{\prime}_{e}(\gamma^{\prime})}{\gamma^{\prime 6} \left(1 -
  \beta_{\gamma^{\prime}} \cos{\psi^{\prime}}\right)^{2}}
\left[e^{\frac{\epsilon^{\prime}_{\rm S} \Gamma_{\rm sh} (1 +
      \beta_{\Gamma_{\rm sh}} \eta^{\prime})}{\Theta \gamma^{\prime 2}
      (1 - \beta_{\gamma^{\prime}} \cos{\psi^{\prime}})}} -
  1\right]^{-1}~.
\end{equation}

\section{\label{paramstud}Parameter Study and Results}

We explore the effects of varying the contribution of the disk, the
BLR, and the DT on the resultant seed photon fields and their
manifestation on the simulated SED and lightcurves of a typical
blazar. This is important for understanding the evolution of the HE
emission of blazars as a function of distance down the jet and thus
gain insight on the location of the observed $\gamma$-ray emission.

\subsection{\label{baseset}Our Baseline Model}

For the purposes of this study, the flux values are calculated for the
frequency range $\nu^{\prime} = (10^{8} - 10^{26})$ Hz and for the
electron energy distribution (EED) range $\gamma^{\prime} = 10 -
10^{8}$, with both ranges divided into 50 grid points. The entire
emission region is divided into 100 slices with 50 slices in the
forward and 50 in the reverse shock regions. The code has been fully
parallelized using the OpenMP interface. This has resulted in
significant speed-up in the time-dependent numerical calculation of
radiative transfer processes in our multi-zone scenario.

Table \ref{basesetlist} shows the values of the base set (run 1)
parameters used to obtain our baseline model. The parameters of this
generic blazar are motivated by a fit to the blazar 3C~279 for
modeling rapid variability on timescales of $\sim$ 1 day. The input
parameters up to $\theta^{*}_{\rm obs}$ have been explained in Paper
1, except for $z_i$ and $z_o$, which refer to the location of inner
and outer shells, respectively, along the jet axis. These values are
used to calculate the point of collision, $z_c$, along the z axis (see
Paper 1 for details), which determines the initial location of the
emission region along the jet axis. According to the model description
given in Paper I, the input parameters for the base set are used to
obtain a value for the BLF of the emission region, which in this case
is $\Gamma_{\rm sh} = 14.9$. The BLF value in turn yields a magnetic
field strength of $B^{\prime} = 2.71$ G and $\gamma^{\prime}_{\rm max}
= 1.26 \times 10^{5}$ for both the forward and reverse emission
regions. On the other hand, the value of $\gamma^{\prime}_{\rm min}$
is numerically obtained to be $8.81 \times 10^{2}$ for the forward and
$1.85 \times 10^{3}$ for the reverse emission regions. Similarly, the
total widths of the forward and reverse emission regions are
analytically obtained to be $\Delta^{\prime}_{\rm fs} = 1.01 \times
10^{16}$ cm and $\Delta^{\prime}_{\rm rs} = 2.12 \times 10^{16}$ cm,
respectively, which consequently yields a shock crossing time for each
of the emission regions of $t^{\prime}_{\rm cr, fs} = 8.93 \times
10^{5}$ s and $t^{\prime}_{\rm cr, rs} = 1.37 \times 10^{6}$ s. In the
observer's frame, this corresponds to the forward shock (FS) spending
$\sim 15$ hours in the forward emission region and the reverse shock
(RS) spending $\sim 23$ hours in the reverse emission region. The
width, and consequently shock crossing time, for each of the emission
regions is set such that it is comparable to the flaring period of our
simulation.

The inner and outer shells collide at a distance of $z_c = 1.01 \times
10^{17}$ cm, making this the starting position of the emission region
along the jet axis. The entire simulation runs for a total of $\sim 5$
days in the observer's frame, during which the emission region moves
beyond the BLR and into the DT, covering a distance of 1.04 pc, in the
AGN frame. For our baseline model, the forward shock exits the forward
emission region within a day in the observer's frame, when the
emission region is located in the cavity of the BLR at $\sim 0.16$
pc. Similarly, the reverse shock exits its region within a day when
the emission region is located within the BLR at $\sim 0.24$ pc. Over
the time scale of our simulation, the BLR energy density,
$u^{\prime}_{\rm ph, BLR}$, changes from $2.82 \times 10^{-2}$ to
$1.18 \times 10^{-5}~erg ~cm^{-3}$, while the DT energy density,
$u^{\prime}_{\rm ph, DT}$, evolves from $3.62 \times 10^{-2}$ to $5.62
\times 10^{-3}~erg ~cm^{-3}$.

\begin{deluxetable}{ccc}
\tabletypesize{\scriptsize}
\tablecaption{Parameter list of run 1 used to obtain the baseline
  model. \label{basesetlist}}
\tablewidth{0pt}
\tablehead{
\colhead{Parameter} & \colhead{Symbol} & \colhead{Value}
}
\startdata
Kinetic Luminosity & $L_w$ & $5 \times 10^{47}$~erg/s\\
Event Duration & $t_w$ & $10^{7}$~s\\
Outer Shell Mass & $M_o$ & $1.531 \times 10^{32}$~g\\
Inner Shell BLF & $\Gamma_i$ & 26.3\\
Outer Shell BLF & $\Gamma_o$ & 10\\
Inner Shell Width & $\Delta_i$ & $6.2 \times 10^{15}$~cm\\
Outer Shell Width & $\Delta_o$ & $7.4 \times 10^{15}$~cm\\
Inner Shell Position & $z_i$ & $7.8 \times 10^{15}$~cm\\
Outer Shell Position & $z_o$ & $1.56 \times 10^{16}$~cm\\
Electron Energy Equipartition Parameter & $\varepsilon^{\prime}_e$ &
$9 \times 10^{-2}$\\
Magnetic Energy Equipartition Parameter & $\varepsilon^{\prime}_B$ &
$2.5 \times 10^{-3}$\\
Fraction of Accelerated Electrons & $\zeta^{\prime}_e$ & $1 \times
10^{-2}$\\
Acceleration Timescale Parameter & $\alpha^{\prime}$ & $2 \times
10^{-5}$\\
Particle Injection Index & $q^{\prime}$ & 4.0\\
Slice/Jet Radius & $R^{\prime}_z$ & $5.44 \times 10^{16}$~cm\\
Observer Frame Observing Angle & $\theta^{*}_{\rm obs}$ & $1.5
\deg$\\
Disk Luminosity & $L_{\rm disk}$ & $8 \times 10^{45}$~erg/s\\
BH Mass & $M_{\rm BH}$ & $2 \times 10^{8} M_{\odot}$\\
Accretion Efficiency & $\eta_{\rm acc}$ & 0.06\\
BLR Luminosity & $L_{\rm BLR}$ & $8 \times 10^{44}$~erg/s\\
BLR inner radius & $R_{\rm in, BLR}$ & $6.17 \times 10^{17}$~cm\\
BLR outer radius & $R_{\rm out, BLR}$ & $1.85 \times 10^{18}$~cm\\
BLR optical depth & $\tau_{\rm BLR}$ & 0.01\\
BLR covering factor & $f_{\rm cov, BLR}$ & 0.03\\
DT inner radius & $R_{\rm in, DT}$ & $3.086 \times 10^{18}$~cm\\
DT outer radius & $R_{\rm out, DT}$ & $6.17 \times 10^{18}$~cm\\
Ldisk fraction & $\xi$ & 0.2\\
DT covering factor & $f_{\rm cov, DT}$ & 0.2\\
Redshift & $Z^{*}$ & 0.538\\
\enddata
\end{deluxetable}

\begin{figure}
\plottwo{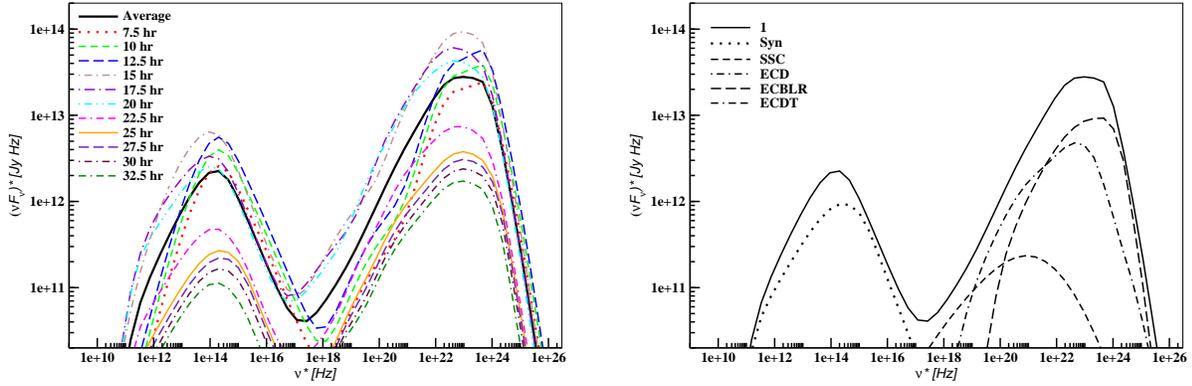}{f10}
\caption{Sample SEDs from the computations. Left: Simulated
  instantaneous spectra of the baseline model of a generic blazar. The
  thick solid black line shows the SED resulting from averaging over
  an integration period of 1 day, corresponding to a rapid flare of
  that duration. The SEDs corresponding to 9 ks and 18 ks have not
  been considered in time-averaging since they correspond to a state
  when the system is still chaotic as it gradually approaches
  equilibrium. Right: Time-averaged SED showing individual radiation
  components as dotted: synchrotron; small-dashed: SSC; dot-dashed:
  ECD, which cannot be seen since its contribution is below $10^{10}$
  Jy Hz for this case; long-dashed: ECBLR; and dot-double dash: ECDT.}
\label{1seds}
\end{figure}

Figure \ref{1seds} shows the instantaneous broadband spectra and the
time-averaged SED from the baseline model. Since we are focussing on
rapid variability in this study, we have evaluated the SED averaged
over an integration time of $\sim$ 1 day. As mentioned in Paper 1,
each instantaneous spectrum shown in Fig. \ref{1seds} corresponds to a
combination of multiple instantaneous SEDs, from both forward and
reverse emission regions, binned over a time period of 9 ks. This was
done to facilitate file management on the computational facility being
used and to be able to compare instantaneous SEDs to X-ray
observations, which have a typical integration time of the same
order. The time-averaged SED is shown by the heavy solid curve on the
left side of Fig. \ref{1seds}, while the right side illustrates
time-averaged radiative components responsible for the total
time-averaged SED. In our framework, although the time-averaged
components dictate the overall profile of the SED and clearly show
which component is responsible for emission in a particular energy
band, they do not exactly match the level of the total time-averaged
SED. This is because, in our model, individual radiative components
are calculated from the emission coefficients rather than from the
actual escaping radiative flux.

The instantaneous SEDs shown in the left hand side of Fig. \ref{1seds}
exhibit the effects of acceleration and cooling on the broadband
spectra of our generic blazar in a time-dependent manner. As the
shocks propagate through the system and energize an increasingly
larger volume of the emitting regions, the overall flux level of the
spectra continues to increase without affecting the location of peak
frequencies for the synchrotron and EC component. Once the emission
from the system reaches its maximum (at $\sim$ 15 hr in this case), by
which time the forward shock has already left the forward emission
region, cooling starts to show its effects on the SEDs, with the
entire broadband spectrum extending to progressively lower frequencies
and the overall flux declining steadily. At later times ($\sim$ 17.5
hr onwards), the emission comes from a comparatively smaller volume of
the emitting region, with the reverse emission region contributing the
most at this time since the reverse shock is still present in the
system. This implies that fresh high-energy electrons, which dominate
the emission at the synchrotron peak, are still being injected into
the system at that time. Consequently, the synchrotron component
after 17.5 hr does not progress to lower frequencies, although the
high-energy component does.

\begin{figure}
\plotone{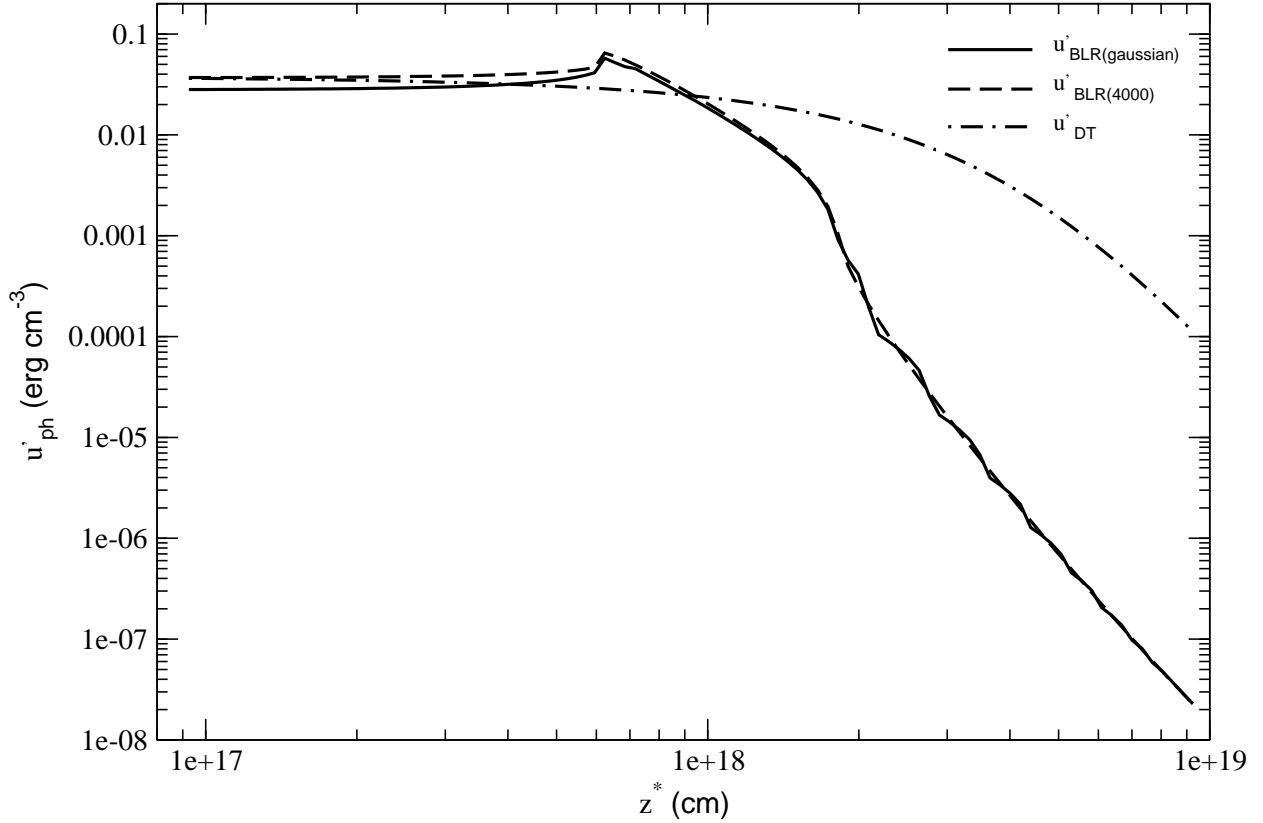}
\caption{Energy density profiles of BLR and DT photon fields for the
  baseline model. The black solid curve denotes BLR energy density
  profile calculated using a Gaussian quadrature method for evaluating
  integrals over BLR angles. The long-dashed curve represents the same
  as mentioned above, but using a linear grid for BLR angles
  consisting of 4000 points. The percent difference between the two
  approaches is 27\%. The dot-dashed curve illustrates the DT energy
  density profile out to a distance of $\sim$ 3 pc.}
\label{1uphcom}
\end{figure}

As can be seen from the right side of Fig. \ref{1seds}, the EC
component peaks in the $\gamma$-ray regime at $\sim 10 \times 10^{23}$
Hz, while the synchrotron component peaks in the near-IR at $\sim 2
\times 10^{14}$ Hz. The transition from synchrotron to high-energy
emission takes place in the X-ray range at $\sim 3 \times 10^{17}$
Hz. As can be seen from Fig. \ref{1seds}, the EC emission of the base
set is dominated by the ECBLR component, which peaks at $\sim 0.4$
GeV. For the flux level (in Jy Hz) considered for our cases, the ECD
component does not contribute to the HE component of this blazar,
while the ECDT component is responsible for the emission in the MeV
range reaching its maximum level at $\sim 200$ MeV. As a result, ECBLR
and ECDT are the two major components that govern the cooling of
electrons/positrons due to EC emission. The derived Compton dominance
factor (CDF), defined as CDF = $\nu F^{*~\rm EC, peak}_{\nu} / \nu
F^{*~\rm syn, peak}_{\nu}$, is 12.4. The spectral hardness (SH) of the
SED can be quantified in terms of the photon spectral index, which is
found to be $\alpha^{*}_{\rm 2-10 keV} = 0.59$ in the X-ray (2 - 10
keV) range and is indicative of a hard SSC-dominated X-ray
spectrum. The Fermi range photon spectral index (calculated at 10 GeV)
is $ \alpha^{*}_{\rm 10 GeV} = 2.75$ and implies a much softer
$\gamma$-ray spectrum. The left side of Fig. \ref{1uphcom} shows a
comparison of total energy density, $u^{\prime}_{\rm ph}$ (in units of
$erg~ cm^{-3}$) due to the BLR and DT photon fields for our baseline
model. As can be seen, energy densities due to the two photon fields
are comparable to each other at sub-pc scales, with the BLR energy
density peaking at $\sim$ 0.21 pc and plummeting beyond this. The DT
energy density takes over at $\sim 0.3$ pc and remains the dominant
contributor to the seed photon field out to $\sim$ 3 pc. The
long-dashed curve in the figure shows a more accurate representation
of the BLR energy density, which was obtained by using a linear grid
for $\mu^{\prime}_{\rm ph, BLR}$ consisting of 4000 points. In order
to save computation time in calculating intensities due to BLR line
and diffuse continuum emission at each of these points, we switched to
the Gaussian quadrature method for evaluating integrals over BLR
angles. The Gaussian grid consisted of only 48 points and resulted in
faster calculations. The percent difference between these two
approaches is 27\% in the beginning, reducing to 12\% at the peak of
the BLR energy density profile. The initial difference of 27\% is not
expected to change or affect our inferences on the dominance of a
particular EC process on the overall profile of SEDs, because this
difference is overshadowed by the amount of boosting BLR photons
receive while entering the jet from the front.

\begin{figure}
\plotone{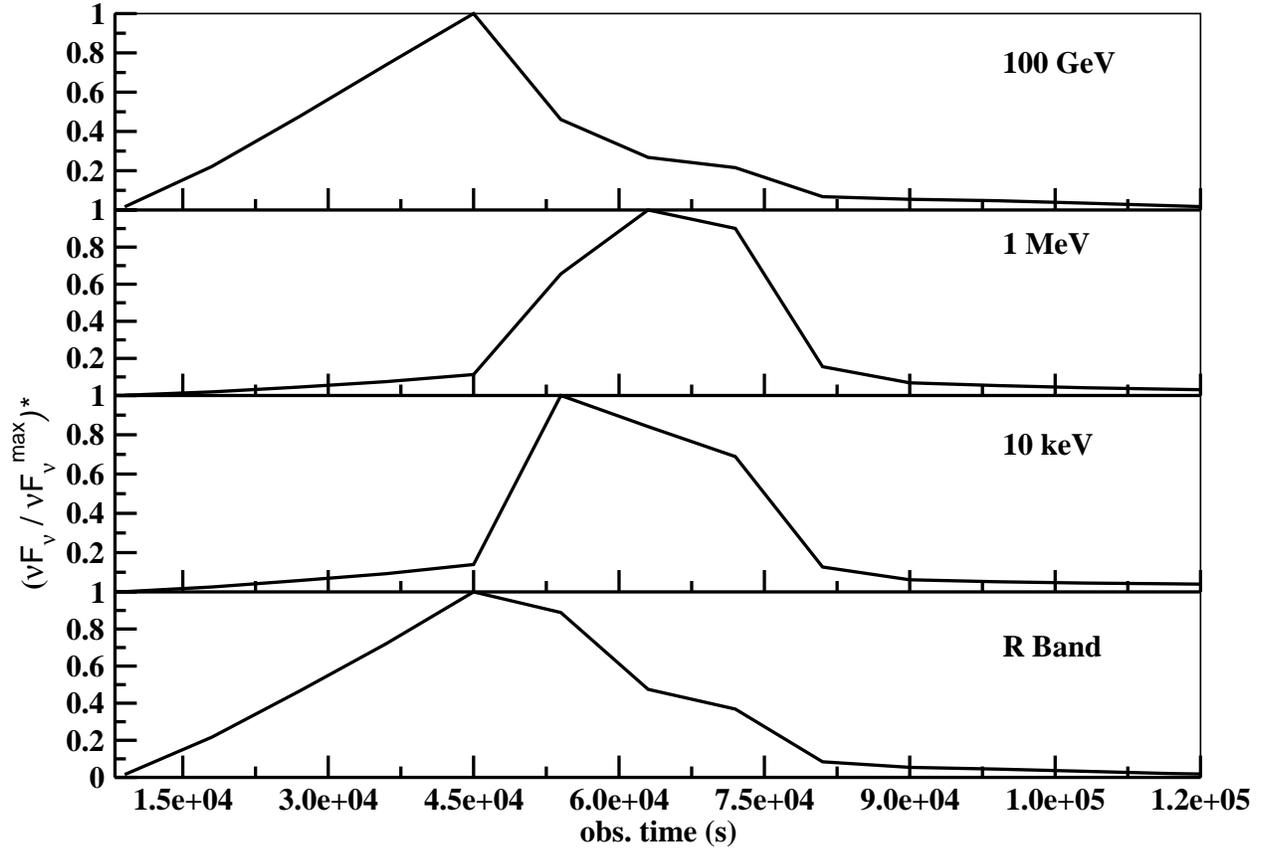}
\caption{Simulated light curves at R band, 10 keV, 1 MeV, and 100 GeV
  energies resulting from our baseline model with parameters listed in
  Table \ref{basesetlist}.}
\label{1lcs}
\end{figure}

Figure \ref{1lcs} shows light curves in the optical (R band), X-ray
(10 keV), HE $\gamma$-ray (1 MeV), and $\gamma$-ray in the upper Fermi
range (100 GeV) spectral regimes from our baseline model for a 1-day
flaring period. As can be seen from the figure, the
synchrotron-dominated optical and EC-dominated HE emission in the 100
GeV energy range are governed by the presence of shocks in the
system. As explained in Paper 1, the respective pulses steadily rise
for as long as the acceleration of particles operates and both reach
their maximum at $t^{*}_{\rm peak}$ = 45 ks, after which they start to
decline rapidly. The Fermi light curve starts to decline sooner and
decays faster than the R-band light curve as long as the FS is present
in the system (until $t^{*}_{\rm FS, exit} \sim$ 53 ks). Once the FS
exits, the Fermi light curve becomes shallower, while the R-band light
curve undergoes a sharper decline, which is marked by a break in the
decaying part of the respective pulse profiles. This is because, as
also mentioned in Paper 1, higher energy electrons are involved in
producing optical synchrotron and 100 GeV EC photons. Such electrons
cool on a timescale shorter than the dynamical timescale within a
particular zone. Thus, once the shocks exit their respective emission
regions and radiative cooling prevails in that region, the optical and
100 GeV pulse rapidly decay. This makes the rising and decaying phases
of the pulse nearly equal and result in a quasi-symmetrical pulse
profile. The X-ray light curve at 10 keV is a result of upscattering
of lower energy (near IR) synchrotron photons by lower energy
electrons and is dominated by the low-energy end of SSC
emission. Since such electrons remain in the system for an extended
period of time, the X-ray light curve peaks later than the optical and
Fermi light curves. At the same time, there is a continued build-up of
late-arriving photons at scattering sites, due to which the pulse
peaks later and exhibits a much more gradual decline, resulting in an
asymmetrical pulse profile \citep{jb2011}. The 1 MeV light curve, on
the other hand, results from the rising part of the ECDT emission,
with some contribution from that of the ECBLR component. This implies
that lower energy electrons are responsible for emission in this
energy range compared to those responsible for the optical and 100 GeV
emission. As a result, the 1 MeV flux is last to peak at $t^{*}_{\rm
  peak}$ = 63 ks. Since the timescale of decay is inversely
proporational to the characteristic energy of the electrons
responsible for the respective emission, the 1 MeV light curve decays
later compared to its 100 GeV counterpart.

\subsection{\label{outcome}Parameter Variation}

Here, we explore the effects of varying physical parameters related to
the EC emission in order to understand their impact on the evolution
of broadband spectra and light curves of our generic blazar. For all
the cases described below, the simulation run time is the same as that
of the baseline model, which is $\sim$ 5 days in the observer's
frame. Table \ref{paramlist} shows the values of each of the
parameters that are varied in the rest of the simulations. We describe
the effects of varying these parameters on the time-averaged SEDs and
light curves with respect to that of the baseline model in sections
\ref{varz} - \ref{varfcovdt}.

\begin{deluxetable}{ccc}
\tabletypesize{\scriptsize}
\tablecaption{Parameter list for other simulations. The corresponding
  values of the baseline model are also listed here for
  reference. \label{paramlist}}
\tablewidth{0pt}
\tablehead{
\colhead{Run \#} & \colhead{Parameter Value} & \colhead{Baseline Model}
}
\startdata
2 & $z_c = 0.05~ R_{\rm in, BLR}$ & $0.16~ R_{\rm in, BLR}$\\
3 & $z_c = 0.81~ R_{\rm in, BLR}$\\
4 & $z_c = 3.10~ R_{\rm in, BLR}$\\
5 & $L_{\rm BLR} = 8 \times 10^{43} erg~s^{-1}$ & $8 \times 10^{44} erg~s^{-1}$\\
6 & $L_{\rm BLR} = 7.2 \times 10^{45} erg~s^{-1}$\\
7 & $f_{\rm cov, DT} = 0.01$ & 0.2\\
8 & $f_{\rm cov, DT} = 0.9$\\
\enddata
\end{deluxetable}

\subsubsection{\label{varz}Variations of $z_{c}$}

\begin{figure}
\plottwo{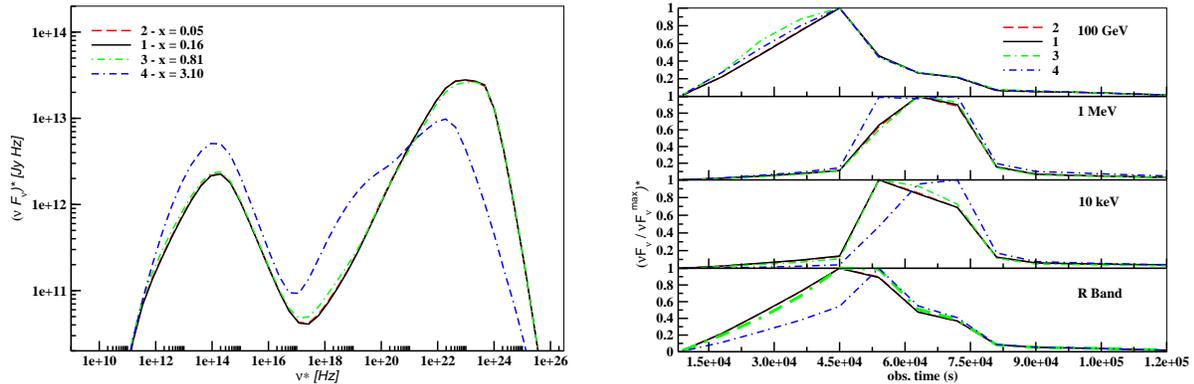}{f15}
\caption{Simulated time-averaged SEDs (left) and light curves (right)
  of a generic blazar from runs 2, 3, and 4, illustrating the effects
  of varying the position, $z_c$, of the emission region along the jet
  axis. The plots are compared against those of run 1 (baseline
  model).}
\label{1234sedlcs}
\end{figure}

Figure \ref{1234sedlcs} shows the impact of changing the starting
position along the jet axis, $z_{c}$, of the emission region on the
time-averaged SEDs and light curves of our baseline model. As
described in \S \ref{baseset}, the point of collision decides the
starting position of the emission region along the z-axis. Changing
this location and understanding its effect on the resultant SED and
light curves is important in comprehending the effect of the AGN
environment as the emission region moves spatially through it.

In the case of run 2, the starting position of the emission region is
at $z_c = 3.08 \times 10^{16}$ cm or $\sim 5 \times 10^{-2}~ R_{\rm
  in, BLR}$. During this run, the emission region moves beyond the
BLR, covering a distance of 1.02 pc, in the AGN frame, similar to that
of run 1. In this case, the FS exits the system when the emission
region is located in the cavity of the BLR, but closer to the central
engine compared to that of run 1, at $z \sim 0.14$ pc or $\sim 0.7~
R_{\rm in, BLR}$. The RS exits the system when the emission region is
located within the BLR at $z \sim 0.21$ pc or $\sim 1.1~ R_{\rm in,
  BLR}$. In the case of run 3, $z_c = 5.02 \times 10^{17} cm$ or $\sim
0.8~ R_{\rm in, BLR}$. Here, the emission region moves slightly deeper
into the DT compared to that of run 1, and ends at a distance of 1.17
pc. In this case the FS exits the system when the emission region is
located within the BLR at $z \sim 0.29$ pc or $\sim 1.5~ R_{\rm in,
  BLR}$ and the RS exits the system when the region is at $z \sim
0.36$ pc or $\sim 1.8~ R_{\rm in, BLR}$. In the case of run 4, the
emission region is placed beyond the BLR at $z_c = 1.91 \times
10^{18}$ cm or $\sim 3~ R_{\rm in, BLR}$. Consequently, the system
evolves until 1.63 pc with the FS exiting the system at $z \sim 0.75$
pc or $\sim 3.7~ R_{\rm in, BLR}$ and the RS at $z \sim 0.82$ pc or
$\sim 4.1~ R_{\rm in, BLR}$.

As can be seen from the left side of Fig. \ref{1234sedlcs}, placing
the emission region closer to the central engine (run 2) or closer to
the inner radius of the BLR (run 3) doesn't change the overall profile
of the time-averaged SED. This is because, just like run 1, in both
cases the high-energy emission continues to be dominated by the ECBLR
process. In other words, the total external radiation energy density,
as received by the emission region in its comoving frame, does not go
through any significant change. The location of peak frequencies for
synchrotron and EC processes, and the location of the transition
frequency from synchrotron to high-energy emission, remains the same
as that of run 1. The CDF value doesn't change between run 1 and 2,
whereas for run 3 the value changes slightly to 11.0. This is because
in this case, since the starting position of the emission region is
located very close to the inner radius of the BLR the amount of
boosting of incoming BLR photons received by the emission region is
slightly lower compared to that of run 1. As a result, the electrons
don't lose as much energy due to the ECBLR process, which slightly
increases the amount of synchrotron emission and brings down the value
of CDF. The SH for the 2-10 keV and Fermi ranges does not change
between runs 1 and 2. For run 3, $\alpha^{*}_{\rm 2-10 keV} = 0.58$
and $\alpha^{*}_{\rm 10 GeV} = 2.74$, representing a minor change in
the SH in these two energy ranges due to reasons mentioned above.

\begin{figure}
\plotone{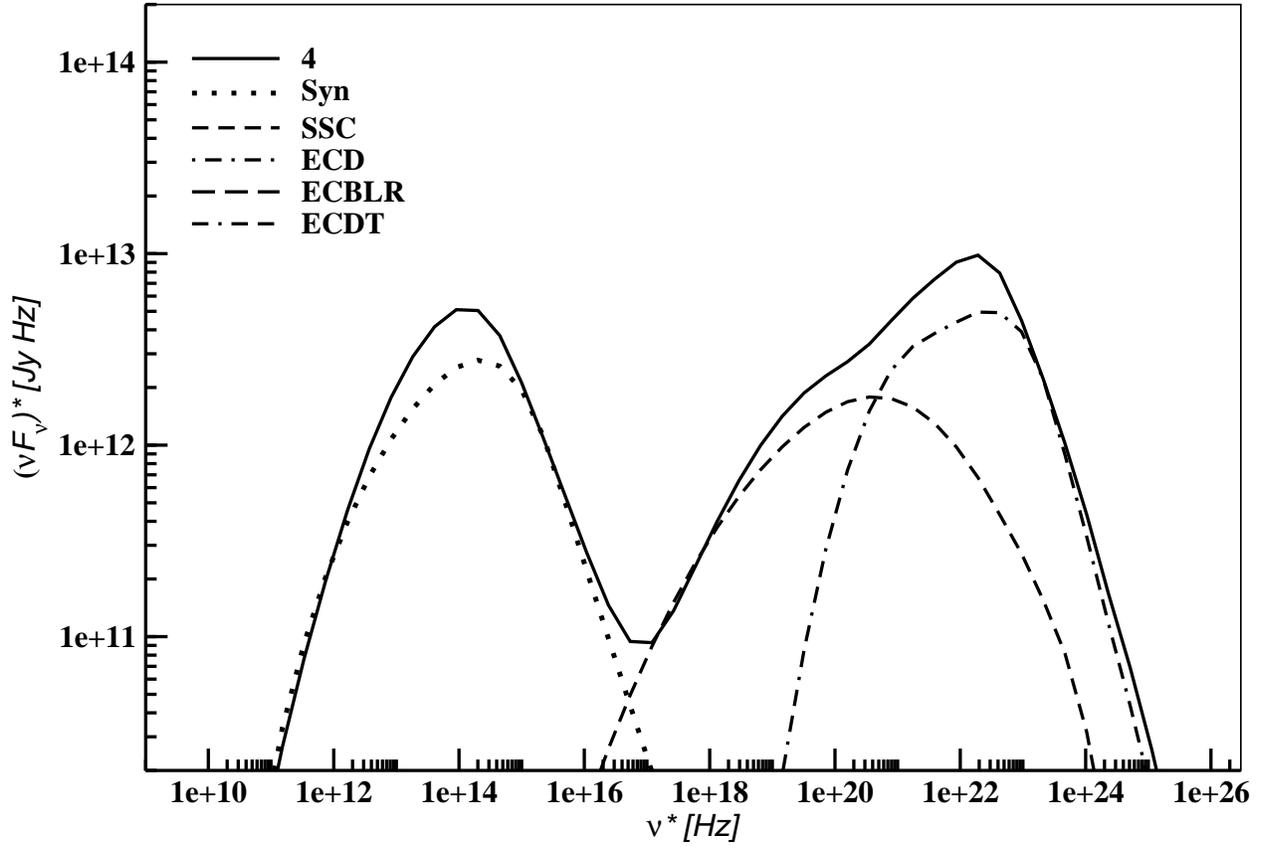}
\caption{Time-averaged SED of run 4 showing the contribution of the
  individual radiation components as dotted: synchrotron;
  small-dashed: SSC; dot-dashed: ECD, which cannot be seen as its
  contribution is below $10^{10}$ Jy Hz; long-dashed: ECBLR, which
  also cannot be seen due to the reason stated above; and dot-double
  dash: ECDT.}
\label{4sedcomps}
\end{figure}

On the other hand, for the case where the emission region is placed
outside the BLR (run 4), the ECBLR is no longer the dominant
process. This is tantamount to the emission region containing a lower
external radiation energy density in its comoving frame. As a result,
the entire high-energy emission is dominated by both ECDT and SSC
processes. This can also be seen in Fig. \ref{4sedcomps}, which
illustrates the individual time-averaged radiation components
responsible for the broadband time-averaged SED of run 4. The ECDT
process is responsible for emission in the MeV - soft GeV range and
peaks at $\sim 120$ MeV, whereas the SSC component makes up the X-ray
to soft-MeV regime and peaks at $\sim 1$ MeV. Since the emission
region is located outside the BLR, it does not receive as much
boosting of BLR photons from the front. The boosting of IR photons
from the dusty torus is also not strong enough at this location, due
to which the overall flux level of the EC component
decreases. Consequently, radiative cooling of electrons due to
synchrotron process becomes stronger, which increases the level of
synchrotron and SSC components in the broadband spectra and brings the
CDF down to 1.93. As a result, the location of peak frequencies for
the synchrotron and high-energy component, as well as the location of
the transition frequency from synchrotron to high-energy emission,
shift to lower frequencies. A by-product of the reduced amount of
radiative cooling due to EC processes is that the SH in the X-ray and
Fermi ranges increases to $\alpha^{*}_{\rm 2-10 keV} = 0.35$ and
$\alpha^{*}_{\rm 10 GeV} = 2.12$, respectively.

The right side of Fig. \ref{1234sedlcs} shows a comparison of light
curves for runs 2, 3, and 4 against those of run 1. As expected, the
light curves of runs 2 and 3 essentially follow the same profile as
those of run 1 due to the continued dominance of the ECBLR process in
both these cases. As a result, the time taken for a pulse to peak and
decline, in a particular energy band, is nearly the same for all three
runs. However in the case of run 3, since the Compton cooling rate due
to the ECBLR process is slightly less compared to that of run 1, for
reasons mentioned above, there is a slight build-up of synchrotron
photons for as long as the shocks are located within the emission
region. This leads to a slightly extended feature in the
synchrotron-dominated R-band pulse compared to that of run 1. On the
other hand, in the case of run 4, due to a much reduced external
Compton cooling rate compared to that of run 1 the entire broadband
spectrum shifts to lower energies. Also, as explained above, this
increases the dominance of synchrotron and SSC emission and allows
them to persist for a longer duration compared to that of run 1. As a
result, both the synchrotron-dominated R-band and SSC-dominated 10 keV
pulses peak later than their run 1 counterparts. However, as explained
in section \ref{baseset}, since the electrons responsible for the
optical synchrotron pulse have higher energy compared to those
responsible for the 10 keV pulse, they last only for as long as the
shocks remain in the system. As a result, even though the R-band pulse
for run 4 attains its peak later, it declines rapidly enough, once the
FS exits the system, to match the decline of its run 1 counterpart,
thereby making the duration of the R-band pulse for run 4
comparatively shorter. The 1 MeV pulse, on the other hand, is
dominated by the peak of SSC emission and some contribution from the
ECDT component. This implies that higher energy electrons than those
emitting the synchrotron-dominated optical pulse are responsible for
emission at this waveband. As a result, the pulse attains its peak
sooner, exhibits a continued gradual buildup of photons, and lasts
longer than its run 1 counterpart. As far as the 100 GeV pulse is
concerned, the pulse profile is similar to that of run 1 due to
reasons explained in section \ref{baseset}.

\subsubsection{\label{varfcovblr}Variations of BLR Covering Factor}

\begin{figure}
\plottwo{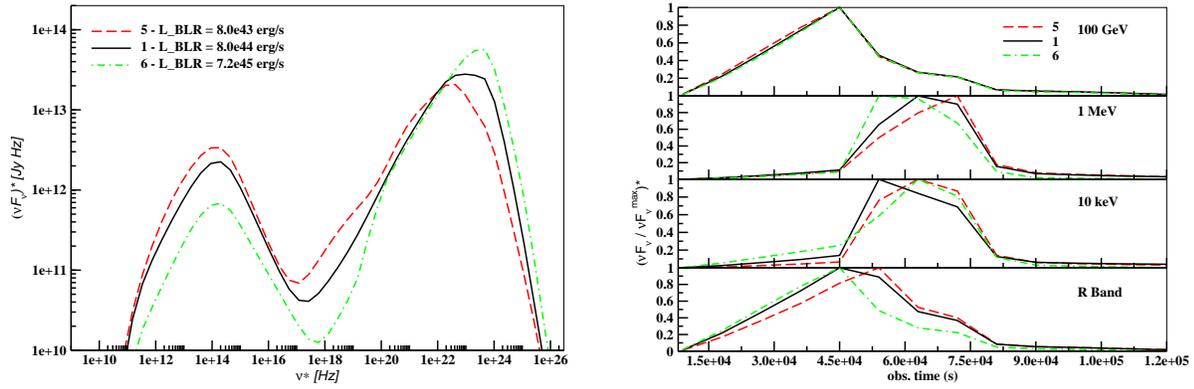}{f18}
\caption{Simulated time-averaged SEDs (left) and light curves (right)
  of a generic blazar from runs 5 \& 6, depicting the effects of
  varying the $L_{\rm BLR}$. The plots are compared against those of
  run 1 (baseline model).}
\label{156sedlcs}
\end{figure}

Figure \ref{156sedlcs} shows the effect of varying the luminosity of
the BLR (runs 5 \& 6) on the time-averaged SED and light curves of the
generic blazar with respect to our baseline model. This is equivalent
to varying the BLR energy density, which directly affects the
lower-energy component of a blazar. Changing $u_{\rm ph, BLR}$ shifts
the partition between the synchrotron and the EC component, which in
turn impacts the synchrotron flux of the SED. In the case of run 5,
$L_{\rm BLR}$ is decreased such that $u^{\prime}_{\rm ph, BLR}$
evolves from $2.82 \times 10^{-3}$ to $1.18 \times 10^{-6}~erg
~cm^{-3}$ over a distance of 1.04 pc. For run 6, $L_{\rm BLR}$ is
increased such that $u^{\prime}_{\rm ph, BLR}$ evolves from $2.54
\times 10^{-1}$ to $1.06 \times 10^{-4}~erg~cm^{-3}$ over 1.04 pc. The
starting position of the emission region for both runs is the same as
that of run 1, with the FS exiting the system in the cavity of the BLR
at $z \sim 0.16$ pc and the RS leaving the system when the emission
region is located within the BLR at $z \sim 0.24$ pc.

\begin{figure}
\plotone{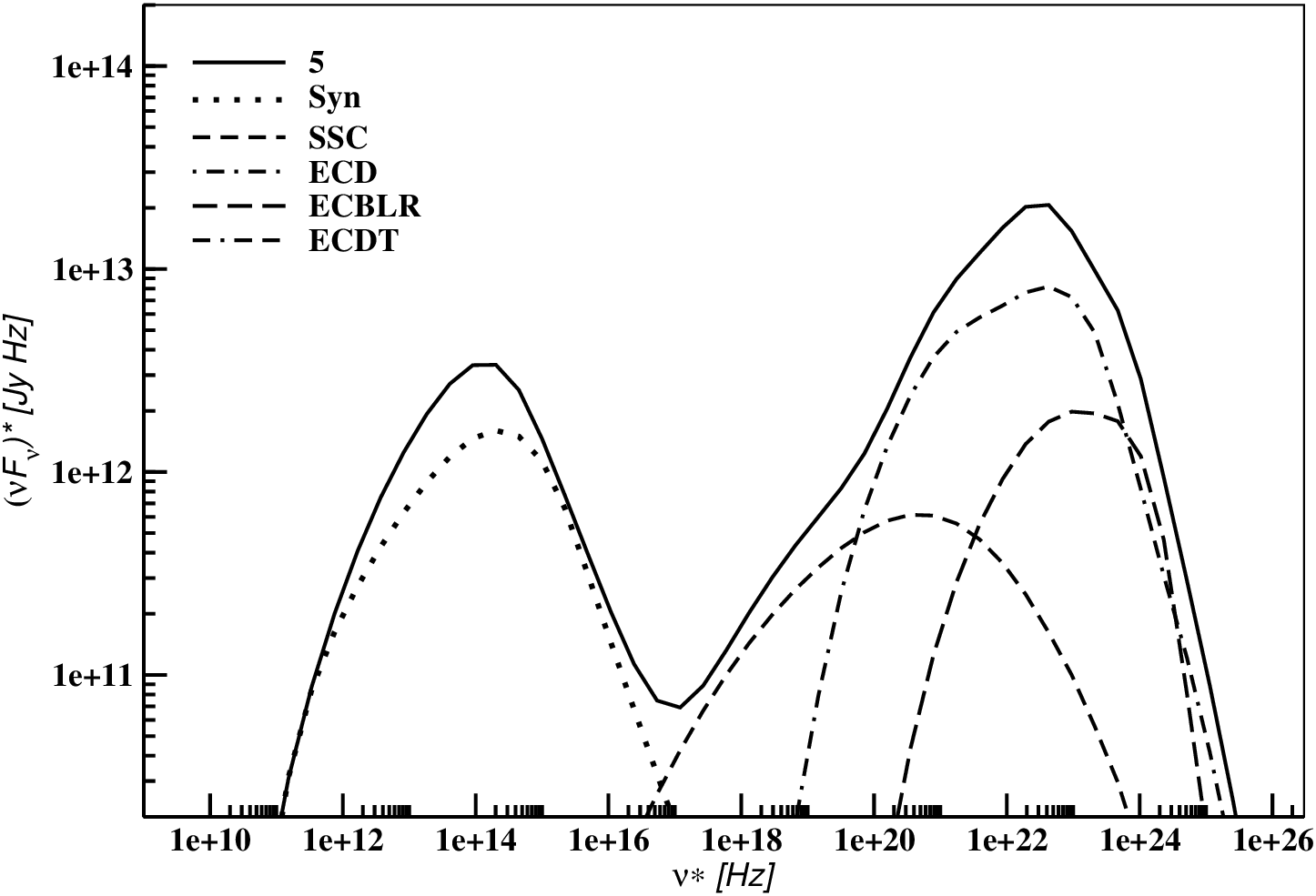}
\caption{Time-averaged SED of run 5 showing the contribution of the
  individual radiation components as dotted: synchrotron;
  small-dashed: SSC; dot-dashed: ECD, which cannot be seen as its
  contribution is below $10^{10}$ Jy Hz; long-dashed: ECBLR, and
  dot-double dash: ECDT.}
\label{5sedcomps}
\end{figure}

As can be seen from the left side of Fig. \ref{156sedlcs}, increasing
the value of $L_{\rm BLR}$ (run 6), is equivalent to increasing the
BLR radiation energy density. This leads to a dominance of the ECBLR
component over the rest of the IC components, which increases the flux
level of the EC component of the spectra and escalates the CDF by six
times in comparison to its run 1 counterpart.  This in turn,
suppresses the SSC component well below the ECDT emission level. The
locations of the transition and peak-EC-component frequencies also
shift to higher values. This is because, in this case, the transition
is no longer between synchrotron-SSC but between synchrotron-ECDT
processes. As a result, the SH in the X-ray regime decreases to
$\alpha^{*}_{\rm 2-10 keV} = 0.70$ because, as can be seen from the
figure, the high-frequency end of the synchrotron component extends
into the 2-10 keV regime, rendering the corresponding spectrum
softer. On the other hand, since the emission at 10 GeV now comes from
a harder part of the spectra, which is closer to the peak of the EC
component compared to that of run 1, the Fermi range spectrum becomes
harder and the SH increases to $\alpha^{*}_{\rm 10 GeV} =
2.50$. Contrary to the above scenario, decreasing the covering factor
of the BLR to 0.01 (run 5), and thereby its $L_{\rm BLR}$, reduces the
dominance of the ECBLR component and increases that of the ECDT
component in producing high-energy emission compared to that of run
1. This can also be seen in Fig. \ref{5sedcomps}, which shows the
individual time-averaged radiation components responsible for the
broadband time-averaged SED of run 5. In addition, due to the reduced
BLR radiation energy density, the flux level of the EC component
decreases compared to that of run 1. This in turn, lowers the CDF to
6.35 and shifts the locations of the transition and peak-EC-component
frequencies leftward. Since the high-frequency end of the synchrotron
compoent no longer extends into the X-ray range in this case, the SH
of the SED in this regime increases to $\alpha^{*}_{\rm 2-10 keV} =
0.49$. On the other hand, the Fermi range SH, with $\alpha^{*}_{\rm 10
  GeV} = 2.61$, is higher than that of run 1 but lower in comparison
to run 6. This is because, in this case, the 10 GeV emission comes
from the softer part of the HE component, which is well below the peak
of the EC component, compared to that of run 6. In addition to this,
there is weaker radiative cooling of particles due to EC processes
compared to that of run 1. As a result, the corresponding Fermi range
spectrum is harder than that of run 1, but softer than that of run 6.

The right side of Fig. \ref{156sedlcs} shows a comparison of light
curves for runs 5 and 6 against those of run 1. The synchrotron R-band
and 10 keV light curves for run 5 peak later compared to those of run
1. This is because, as explained above, a lower $u^{\prime}_{\rm ph,
  BLR}$ results in a reduced Compton cooling rate, which in turn
affects the synchrotron flux and makes the synchrotron-dominated pulse
last longer. As a result, the R-band light curve continues to rise for
as long as the FS is in the system. But, as explained in section
\ref{varz} for the case of run 4, since higher energy electrons are
responsible for the optical synchrotron pulse than those for the 10
keV pulse, the optical pulse declines quickly enough to match the
declining pulse profile of that of run 1, once the FS shock crosses
the system at $t^{*}_{\rm FS, ~exit} \sim$ 53 ks. The 1 MeV pulse, in
the case of run 5, is dominated by the rising but softer part of the
ECDT component. This implies that lower energy particles are
responsible for producing this pulse, compared to those of run 1,
thereby causing it to peak later. On the other hand, in the case of
run 6, the R-band pulse exhibits exactly the opposite effect to that
of run 5 due to an increased Compton cooling rate owing to a higher
value of $u^{\prime}_{\rm ph, BLR}$. As a result, it declines much
more rapidly and lasts for a shorter duration. The 10 keV pulse, in
this case, is dominated by the low-frequency end of the ECDT
process. As a result, the corresponding spectrum is softer because
low-energy electrons are responsible for producing this pulse. This is
why the pulse peaks later than that of run 1, but at the same time as
that of run 5. The 1 MeV pulse, on the other hand, is dominated by the
rising part of the ECBLR component and is due to higher energy
electrons in the region, which is why the pulse starts to peak sooner,
undergoes a gradual buildup, and declines faster compared to that of
run 1. As far as the 100 GeV pulse for runs 5 and 6 is concerned, it
follows the same profile as that of run 1 because of reasons explained
in section \ref{baseset}.

\subsubsection{\label{varfcovdt}Variations of DT Covering Factor}

\begin{figure}
\plottwo{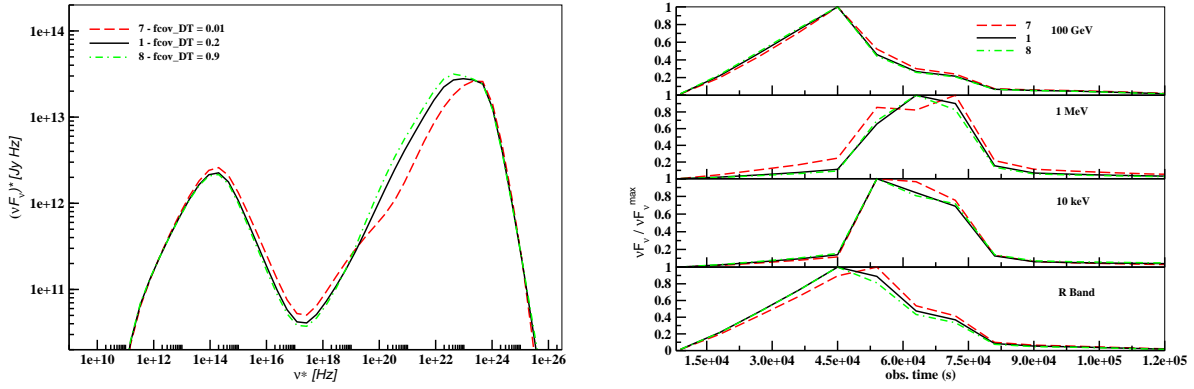}{f21}
\caption{Simulated time-averaged SEDs (left) and light curves (right)
  of a generic blazar from runs 7 \& 8 showing the effects of changing
  the covering factor of the DT and consequently $R_{\rm out, DT}$
  according to Eqs. (\ref{ldteqn}) and (\ref{areadteqn}).}
\label{178sedlcs}
\end{figure}

The impact of decreasing (run 7) or increasing (run 8) the covering
factor of the DT on the time-averaged SED and light curves of a
generic blazar is shown in Fig. \ref{178sedlcs}. This is equivalent to
varying the DT energy density, which again affects the lower-energy
component of a blazar and the location of the subsequent partition
between the synchrotron and EC components. In the case of run 7,
$f_{\rm cov, DT}$ is decreased such that $u^{\prime}_{\rm ph, DT}$
evolves from $5.25 \times 10^{-3}$ to $4.14 \times 10^{-3}~erg
~cm^{-3}$ over a distance of 1.04 pc. For run 8, $f_{\rm cov, DT}$ is
increased such that $u^{\prime}_{\rm ph, DT}$ evolves from $5.72
\times 10^{-2}$ to $3.76 \times 10^{-3} ~erg~cm^{-3}$ over 1.04
pc. The starting position of the emission region for both runs is the
same as that of run 1 with the FS exiting the system in the cavity of
the BLR at $z \sim 0.16$ pc and the RS leaving the system when the
emission region is located within the BLR at $z \sim 0.24$ pc.

\begin{figure}
\plotone{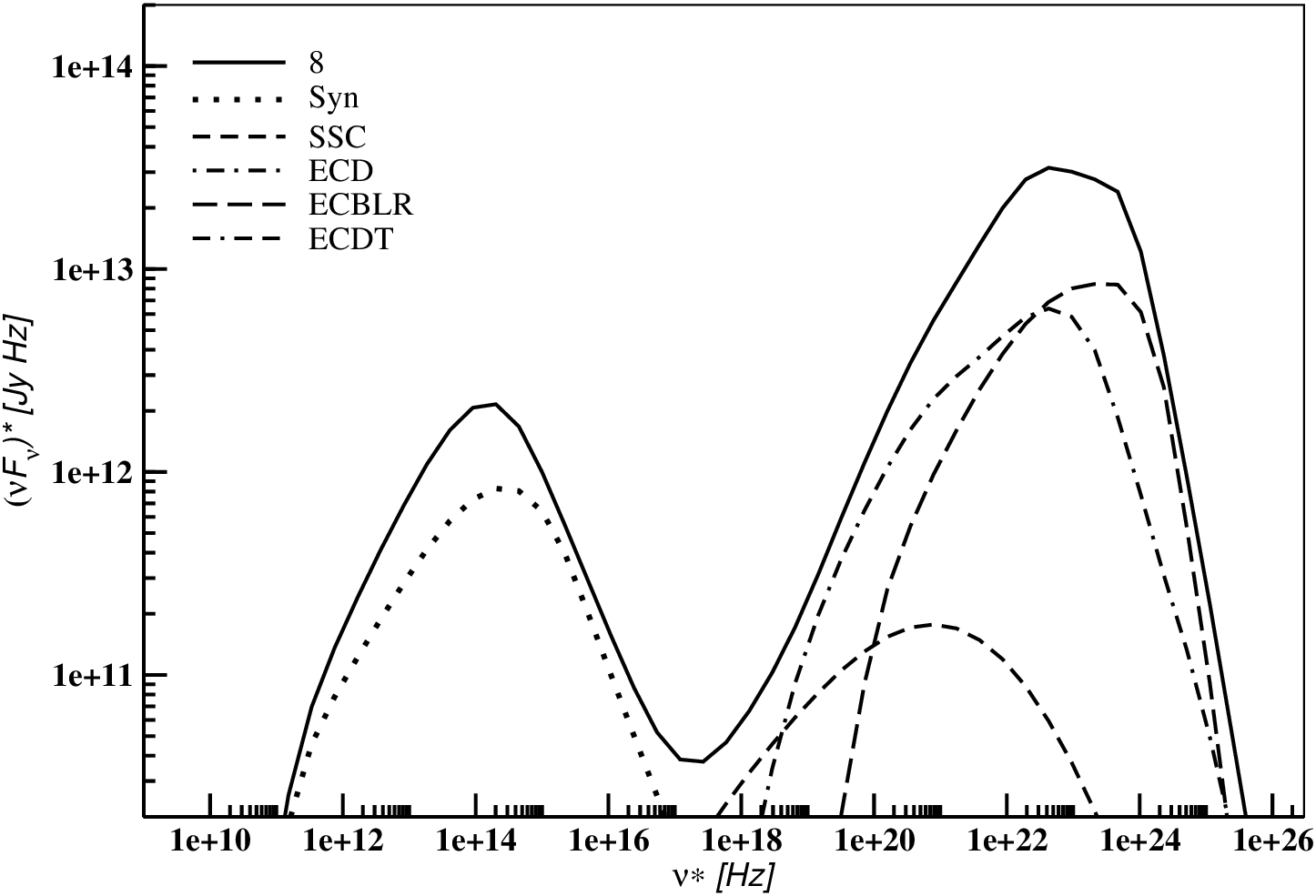}
\caption{Time-averaged SED of run 8 showing the contribution of the
  individual radiation components as dotted: synchrotron;
  small-dashed: SSC; dot-dashed: ECD, which cannot be seen as its
  contribution is below $10^{10}$ Jy Hz; long-dashed: ECBLR, and
  dot-double dash: ECDT.}
\label{8sedcomps}
\end{figure}

Changing $f_{\rm cov, DT}$ also requires changing $R_{\rm out, DT}$,
according to Eqs. (\ref{ldteqn}) and (\ref{areadteqn}), in order to
keep the rest of the input parameters related to the ECDT emission the
same. Decreasing $f_{\rm cov, DT}$ to 0.01 and increasing the size of
the torus such that $R_{\rm out, DT} = 2.37 \times 10^{19}$ cm (run 7)
dilutes the intensity of the ECDT emission. As a result, the
contribution of the ECDT component to the EC emission decreases, which
makes ECBLR the dominant EC component but lowers the overall EC flux
level. Hence, the synchrotron emission rises, slightly decreasing the
CDF to 10.1. The radiative cooling of particles due to EC emission
decreases. This increases the SSC contribution in the X-ray regime,
which consequently increases the SH such that $\alpha^{*}_{\rm 2-10
  keV} = 0.57$. On the other hand, the effects of cooling due to the
ECBLR component increase in the Fermi range, which decrease the SH at
10 GeV such that $\alpha^{*}_{\rm 10 GeV} = 2.77$. The peak
EC-component frequency shifts to higher frequencies due to the
dominance of the ECBLR component in this case. Contrary to the above
scenario, increasing $f_{\rm cov, DT}$ to 0.9 and shrinking the DT
such that $R_{\rm out, DT} = 3.95 \times 10^{18}$ cm (run 8) enhances
the emission of the torus as received by the emission region. This
makes the contribution of the ECDT component comparable to that of the
ECBLR emission, as can be seen from Fig. \ref{8sedcomps}. This has the
opposite effect to that of run 7, although the overall change in the
SED, in this case, is not as significant and closely follows that of
run 1. To summarize the impact, the CDF increases slightly to 14.6 and
the peak EC-component frequency shifts slightly to lower
frequencies. The SH in the X-ray regime decreases slightly compared to
that of run 1, while the SH at 10 GeV increases slightly such that
$\alpha^{*}_{\rm 10 GeV} = 2.73$. This is because, in this case, the
contribution of the ECBLR component decreases slightly compared to
that of run 1, which implies comparatively less cooling and the
presence of more higher-energy electrons, which gives rise to a harder
spectra at that energy range.

The right side of Fig. \ref{178sedlcs} compares light curves for runs
7, and 8 against those of run 1. As can be seen from the figure, the
light curve profiles of run 8 are similar to those of run 1 because
the overall impact of increasing the covering factor of the DT, while
keeping other parameters related to EC emission the same, is not very
significant. On the other hand, in the case of run 7, the
synchrotron-dominated R-band pulse lasts for a slightly longer
duration due to a decreased Compton cooling rate and consequently an
increased amount of synchrotron emission. As a result, it peaks later
than that of run 1 but lasts only for as long as the FS is inside the
emission region and declines afterwards due to the reasons given in
section \ref{baseset}. At 10 keV, the pulse of run 7 peaks at the same
time and has a similar profile as that of run 1, but lasts a bit
longer due to an increased contribution from the SSC emission. On the
other hand, unlike the case for run 1, the 1 MeV emission in run 7 is
dominated by the rising part of the ECBLR component involving
comparatively lower-energy electrons. As a result, the corresponding
pulse peaks later before declining along with those of runs 1 and
8. We would like to point out here that the 1 MeV light curve for run
7 has really only one peak. The slight depression seen at around
63000s is more of a numerical artifact resulting due to the
not-so-high resolution of the electron and photon energy grid. As is
seen in the other cases, the 100 GeV pulse for runs 7 and 8 follows
the same profile as that of run 1 because of reasons explained in
section \ref{baseset}.

\section{\label{disco}Discussion and Conclusions}

We have extended the scope of the multi-zone time-dependent leptonic
jet model, with radiation feedback scheme, in the internal shock
scenario, of \cite{jb2011} (known for short as the MUlti ZOne
Radiation Feedback, MUZORF, model) to include the EC component by
considering anisotropic target radiation fields. The current model is
now applicable to a broader class of blazars, including FSRQs, LBLs,
and IBLs, and allows us to distinguish and quantify the contribution
of each of the seed photon fields in producing high-energy emission of
blazars out to $\sim$ pc scales.

In our approach, we consider three sources of seed photon fields,
namely the accretion disk, the BLR, and the DT. We let the system
evolve to beyond the BLR and into the DT, and calculate the
corresponding distance- and time-dependent EC emission according to
the prescription described in sections \ref{disksec} to
\ref{dtsec}. In our current work, we have not considered detailed
spectral fitting and analysis to compare against the multiwavelength
SEDs and light curves of blazars available through campaigns involving
the {\it{Fermi Gamma Ray Space Telescope}}, which we leave to future
work. Instead, we have carried out a parameter study of rapid
variability, on timescales of $\sim$ 1 day, between optical, X-ray,
and $\gamma$-ray energies. We point out that the MUZORF model is
applicable to EC emission up to only a few parsecs. This is because
currently it does not include adiabatic cooling, which is relevant for
evolving the system out to several parsecs. This is a work in progress
that we plan to address in the future.

We have carried out a parameter study to understand the effects of
varying input parameters relevant to the EC emission on the dynamic
evolution of the SED and light curves of a generic blazar. A number of
input parameters, such as the location of the emission region along
the jet axis ($z_{c}$), the covering factor of the BLR ($f_{\rm cov,
  BLR}$), and that of the dusty torus ($f_{\rm cov, DT}$) were
varied. The goal of the study was to understand the dependence of the
contribution of each of the target photon field components on such
parameters in producing HE emission of blazars. Anisotropic target
radiation fields were considered from each of the seed photon field
components in order to enable the system to evolve beyond the BLR and
allow the contribution of each of the three components to be
incorporated as accurately as possible. As has been demonstrated in \S
\ref{dtsec}, the range of angles covered by incoming DT photons, in
the anisotropic case, is quite narrow. For the cases considered here,
the DT played an important role in contributing toward $\gamma$-ray
emission in the MeV range, while the BLR contributed mostly in the
high MeV - low GeV range. For the value of $L_{\rm disk}$ and $z_{c}$
considered in all of our cases, the emission due to the ECD process
was found to be negligible for the level of fluxes considered
here. This is because of the amount of de-boosting of disk photons
when entering the jet from behind or from the side at those
positions. Similar evolution of the ECD emission with $z_{c}$ was also
found by \cite{de2009}.

For all cases, the 100 GeV light curves always led the X-rays and MeV
emission, while for some cases they either led or peaked at the same
time as the optical light curves, depending on the parameter that was
varied for those simulations. More symmetric pulse profiles were
obtained for higher frequency light curves than those for lower
frequency ones. Similar behavior of pulse profiles were also found by
\citet[]{sm2005, bd2010, jb2011}. It was demonstrated, that for the
case where the emission region was placed beyond the BLR, the HE
emission was primarily due to ECDT and SSC processes. This can be
extrapolated to conclude that as the distance of the emitting plasma
from the central engine increases, SSC starts to dominate over EC
emission due to a substantially reduced amount of Doppler boosting of
incoming seed photon fields. The EC losses then become a fraction of
the synchrotron losses, and SSC flux exceeds the EC flux
\citep{sm2005}. Similar results are achieved if the energy density of
the BLR or DT radiation is reduced, which also introduces an optical
lag relative to the HE $\gamma$-ray variations \citep{bd2010}. In
addition to these input parameters, another factor that plays an
important role in deciding the duration of HE pulses and the range of
the location of $\gamma$-ray emission is the time of exit of shocks
from their respective emitting regions. This is because, as discussed
in \S \ref{paramstud}, $\gamma$-ray emission is produced by highly
energetic leptons, which, in turn, are produced in the system for as
long as the shocks are present and accelerating particles to such high
energies. Therefore, analyzing the effects of the combination of the
factors discussed above on the evolution of the system is crucial to
understanding the origin of $\gamma$-ray emission and its relationship
with emission at lower frequencies.

Understanding of the sources of seed photons is imperative in order to
comprehend the HE emission of blazars in general. At the same time, it
is important to realize that at distances of several parsecs from the
central engine, the target photon field from these components is quite
weak. In that case, these three conventional sources might not explain
the high $\gamma$-ray luminosity and correlation between $\gamma$-ray
and radio events that have been observed at such distances
\citep{jo2013, ma2012, ag2011}. In such a scenario, considering
possible new sources of seed photon fields or modifying the placement
of the existing ones becomes important. The latter might include, (1)
a variable location of the $\gamma$-ray emitting zone \citep{sp2011},
(2) an outflowing BLR serving as a source of external seed photons at
parsec scales \citep{lt2011}, (3) the presence of some stray BLR
material surrounding the radio core that could produce required
density of target photons at those distances \citep{lt2013}, or (4) an
additional but internal source of seed photons, such as a Mach disk,
in tandem with DT photon field \citep{ma2013}. Such possibilities
should be explored in producing the $\gamma$-ray luminosity at those
length scales.

\acknowledgements 

We thank Dr. Jack Steiner and Dr. Justin Finke for useful discussions
and comments. We acknowledge Mr. Karthikeyan Karunanidhi for his help
with computational work. We thank the referee for his/her helpful
comments. The effort of M.J. and A. P. M. in this project has been
supported by NASA through Fermi grants NNX11AQ03G and
NNX12AO59G. M.B. acknowledges support by the South African Research
Chairs Initiative (grant no. 64789) of the Department of Science and
Technology and the National Research Foundation of South Africa.

\appendix
\section{\label{loseqns}Lines of Sight of Incoming BLR Photons}

Here we calculate the respective path lengths of incoming BLR photons
for Pos. 1, 2, and 3 of the emission region, as described in \S
\ref{blrsec}. These path lengths are used to integrate the emission
coefficient to obtain the intensity of incoming photons (in units of
$erg~ s^{-1} cm^{-2} ster^{-1}$) as a function of distance, z, along
the jet axis and angle $\theta$. All quantities in this section refer
to the lab frame. Referring to Fig. \ref{blrgeom}, the quantity $s$ is
considered as $l$ in this section. Then, the values of $l_{\rm min}$
and $l_{\rm max}$ of an incoming BLR photon when the emission region
is located at Pos. 1 are obtained using the cosine law of angles:

\begin{equation}
\label{pos1eqn}
r^{2} = z^{2} + l^{2} - 2zl \cos{\theta}~.
\end{equation}

Given $R_{\rm in}$ and $R_{\rm out}$, we obtain the respective
pathlength limits, $l_{\rm min}$ and $l_{\rm max}$, as

\begin{eqnarray}
\label{pos1leqn}
l_{\rm min} = z \cos{\theta} + z \sqrt{\cos^{2}{\theta} +
  \left(\frac{R_{\rm in}}{z}\right)^{2} - 1}~, \nonumber\\
l_{\rm max} = z \cos{\theta} + z \sqrt{\cos^{2}{\theta} +
  \left(\frac{R_{\rm out}}{z} \right)^{2} - 1}~, &
  ~\textrm{and} \nonumber\\
I(z, \theta) = \int\limits_{l_{\rm min}}^{l_{\rm max}} j(r) dl~,
\end{eqnarray}
where $j(r)$ is the emission coefficient at the radial distance $r$,
as described in \S\ref{blrsec} and \cite{lb2006}. Similarly, when the
emission region is located at Pos. 2, the limits of integration can be
calculated using the method described above, while keeping in mind
that, for this case, $z > R_{\rm in}$. As a result, we obtain

\begin{eqnarray}
\label{pos2eqn}
l_{\rm min, min} = z \cos{\theta} - z \sqrt{\cos^{2}{\theta} +
  \left(\frac{R_{\rm in}}{z}\right)^{2} - 1}, \nonumber\\
l_{\rm min, max} = z \cos{\theta} + z \sqrt{\cos^{2}{\theta} +
  \left(\frac{R_{\rm in}}{z}\right)^{2} - 1}, &
\textrm{and} \nonumber\\
l_{\rm max} = z \cos{\theta} + z \sqrt{\cos^{2}{\theta} +
  \left(\frac{R_{\rm out}}{z}\right)^{2} - 1}, & \textrm{such
  that} \nonumber\\
I(z, \theta) = \int\limits_{0}^{l_{\rm min, min}} j(r) dl +
\int\limits_{l_{\rm min, max}}^{l_{\rm max}} j(r) dl~.
\end{eqnarray}
For $\theta \geq \theta_{\rm cr}$, where $\theta_{\rm cr} =
\cos^{-1}(\frac{l_{\rm cr}}{z})$ and $l_{\rm cr} = \sqrt{z^{2} -
  R^{2}_{\rm in}}$, there is no contribution from lines of sight
through the cavity of the BLR. In this case, the intensity calculation
reduces to $I(z, \theta) = \int\limits_{0}^{l_{\rm max}} j(r) dl$,
where $l_{\rm max}$ is given in Eq. (\ref{pos2eqn}).

For the case where the emission region is located at Pos. 3, critical
angles and their corresponding path lengths to the outer boundary of
the inner and outer circles, respectively, are given by

\begin{eqnarray}
\label{critangleeqn}
\theta_{\rm in, cr} = \cos^{-1}{(\frac{l_{\rm in, cr}}{z})}, \nonumber\\
l_{\rm in, cr} = \sqrt{z^{2} - R^{2}_{\rm in}}, &
\textrm{and} \nonumber\\
\theta_{\rm out, cr} = \cos^{-1}{(\frac{l_{\rm out, cr}}{z})}, \nonumber\\
l_{\rm out, cr} = \sqrt{z^{2} - R^{2}_{\rm out}}~.
\end{eqnarray}

Thus, if $\theta < \theta_{\rm in, cr}$, the intensity would be
obtained according to

\begin{eqnarray}
\label{pos3eqn1}
l_{\rm min, min} = z \cos{\theta} - z \sqrt{\cos^{2}{\theta} +
  \left(\frac{R_{\rm in}}{z}\right)^{2} - 1}, \nonumber\\
l_{\rm min, max} = z \cos{\theta} + z \sqrt{\cos^{2}{\theta} +
  \left(\frac{R_{\rm in}}{z}\right)^{2} - 1}, &
\textrm{and} \nonumber\\
l_{\rm max, min} = z \cos{\theta} - z \sqrt{\cos^{2}{\theta} +
  \left(\frac{R_{\rm out}}{z}\right)^{2} - 1}, \nonumber\\
l_{\rm max, max} = z \cos{\theta} + z \sqrt{\cos^{2}{\theta} +
  \left(\frac{R_{\rm out}}{z}\right)^{2} - 1}, & \textrm{such
  that} \nonumber\\
I(z, \theta) = \int\limits_{l_{\rm max, min}}^{l_{\rm min, min}} j(r) dl +
\int\limits_{l_{\rm min, max}}^{l_{\rm max, max}} j(r) dl~.
\end{eqnarray}

If $\theta_{\rm in, cr} \leq \theta < \theta_{\rm out, cr}$, the
intensity is given by

\begin{eqnarray}
\label{pos3eqn2}
l_{\rm max, min} = z \cos{\theta} - z \sqrt{\cos^{2}{\theta} +
  \left(\frac{R_{\rm out}}{z}\right)^{2} - 1}, \nonumber\\
l_{\rm max, max} = z \cos{\theta} + z \sqrt{\cos^{2}{\theta} +
  \left(\frac{R_{\rm out}}{z}\right)^{2} - 1}, & \textrm{such
  that} \nonumber\\
I(z, \theta) = \int\limits_{l_{\rm max, min}}^{l_{\rm max, max}} j(r) dl~,
\end{eqnarray}


\begin{thebibliography}{}

\bibitem[Abdo et al.(2010)]{ab2010}
Abdo, A. A., et al., 2010, ApJ, 716, 30

\bibitem[Agudo et al.(2011)]{ag2011}
Agudo, I., et al., 2011, ApJL, 735, 10

\bibitem[Albert et al.(2008)]{al2008}
Albert, J., et al., 2008, Science, 320, 1752

\bibitem[Aleksi\'{c} et al.(2012)]{aj2012}
Aleksi\'{c}, J., et al., 2012, A \& A, 542, 100

\bibitem[Blandford \& Levinson(1995)]{bl1995}
Blandford, R. D., \& Levinson, A., 1995, ApJ, 441, 79

\bibitem[{B{\l}a{\.z}ejowski} et al.(2000)]{bl2000}
{B{\l}a{\.z}ejowski}, M., Sikora, M., Moderski, R., \& Madejski, G. M., 2000, ApJ, 545, 107

\bibitem[Bloom \& Marscher(1996)]{bm1996}
Bloom, S. D., \& Marscher, A. P., 1996, ApJ, 461, 657

\bibitem[B\"{o}ttcher(2012)]{bm2012}
B\"{o}ttcher, M., 2012, Fermi \& Jansky Proceedings, arXiv:1205.0539

\bibitem[B\"{o}ttcher \& Bloom(2000)]{bb2000}
B\"{o}ttcher, M., \& Bloom, S. D., 2000, AJ, 119, 469

\bibitem[B\"{o}ttcher \& Dermer(2010)]{bd2010}
B\"{o}ttcher, M., \& Dermer, C. D., 2010, ApJ, 711, 445

\bibitem[B\"{o}ttcher, Mause, \& Schlickeiser(1997)]{bms1997}
B\"{o}ttcher, M., Mause, H., \& Schlickeiser, R., 1997, A\&A, 324, 395

\bibitem[B\"{o}ttcher \& Reimer(2004)]{br2004}
B\"{o}ttcher, M., \& Reimer, A., 2004, ApJ, 609, 576

\bibitem[B\"{o}ttcher, Reimer, \& Marscher(2009)]{brm2009}
B\"{o}ttcher, M., Reimer, A., \& Marscher, A. P., 2009, ApJ, 703, 1168

\bibitem[Chiaberge \& Ghisellini(1999)]{cg1999}
Chiaberge, M., \& Ghisellini, G., 1999, MNRAS, 306, 551

\bibitem[Collmar et al.(2010)]{co2010}
Collmar, W., et al., 2010, A \& A, 522, 66

\bibitem[Dermer et al.(2009)]{de2009}
Dermer, C. D., Finke, J. D., Krug, H., \& B\"{o}ttcher, M., 2009, ApJ, 696, 32

\bibitem[Dermer \& Menon(2009)]{dm2009} 
Dermer, C. D., \& Menon, G., 2009, High Energy Radiation from Black
Holes: \\ Gamma Rays, Cosmic Rays, and Neutrinos, Princeton University
Press

\bibitem[Dermer \& Schlickeiser(2002)]{ds2002}
Dermer, C. D., \& Schlickeiser, R., 2002, ApJ, 575, 667 

\bibitem[Dermer \& Schlickeiser(1994)]{ds1994}
Dermer, C. D., \& Schlickeiser, R., 1994, ApJS, 90, 945 

\bibitem[Dermer \& Schlickeiser(1993)]{ds1993}
Dermer, C. D., \& Schlickeiser, R., 1993, ApJ, 416, 458 

\bibitem[Dermer, Sturner, \& Schlickeiser(1997)]{dss1997}
Dermer, C. D., Sturner, S. J., \& Schlickeiser, R., 1997, ApJS, 109, 103

\bibitem[Donea \& Protheroe(2003)]{dp2003}
Donea, A-C., \& Protheroe, R. J., 2003, Astroparticle Physics, 18, 377

\bibitem[Finke, Dermer, \& B\"{o}ttcher(2008)]{fdb2008}
Finke, J. D., Dermer, C. D., \& B\"{o}ttcher, M, 2008, ApJ, 686, 181

\bibitem[Francis et al.(1991)]{fr1991}
Francis, P., et al., 1991, ApJ, 373, 465

\bibitem[Gaskell, Shields, \& Wampler(1981)]{gsw1981}
Gaskell, C. M., Shields, G. A., \& Wampler, E. J., 1981, ApJ, 249, 443

\bibitem[Ghisellini \& Madau(1996)]{gm1996}
Ghisellini, G., \& Madau, P., 1996, MNRAS, 280, 67

\bibitem[Ghisellini \& Tavecchio(2009)]{gt2009}
Ghisellini, G., \& Tavecchio, F., 2009, MNRAS, 397, 985

\bibitem[Jorstad et al.(2013)]{jo2013}
Jorstad, S. G., et al., 2013, ApJ, 773, 147

\bibitem[Jorstad et al.(2005)]{jo2005}
Jorstad, S. G., et al., 2005, AJ, 130, 1418

\bibitem[Joshi \& B\"{o}ttcher(2011)]{jb2011}
Joshi, M., \& B\"{o}ttcher, M., 2011, ApJ, 727, 21

\bibitem[Kaspi \& Netzer(1999)]{kn1999}
Kaspi, S., \& Netzer, H., 1999, ApJ, 524, 71

\bibitem[Kataoka et al.(1999)]{ka1999}
Kataoka, J., et al., 1999, ApJ, 514, 138

\bibitem[Kusunose \& Takahara(2005)]{kt2005}
Kusunose, M., \& Takahara, F., 2005, ApJ, 621, 285

\bibitem[Le{\'o}n-Tavares et al.(2013)]{lt2013}
Le{\'o}n-Tavares, J., et al., 2013, ApJL, 763, 36

\bibitem[Le{\'o}n-Tavares et al.(2012)]{lt2012}
Le{\'o}n-Tavares, J., et al., 2012, ApJ, 754, 23

\bibitem[Le{\'o}n-Tavares et al.(2011)]{lt2011}
Le{\'o}n-Tavares, J., et al., 2011, A\&A, 532, 146

\bibitem[Lindfors, Valtaoja, \& T{\"u}rler(2005)]{lvt2005}
Lindfors, E. J., Valtaoja, E., \& T{\"u}rler, M., 2005, A\&A, 440, 845

\bibitem[Liu \& Bai(2006)]{lb2006}
Liu, H. T., \& Bai, J. M., 2006, ApJ, 653, 1089

\bibitem[Malmrose et al.(2011)]{ma2011}
Malmrose, M., et al., 2011, ApJ, 732, 116

\bibitem[Marscher et al.(2012)]{ma2012}
Marscher, A. P., et al., 2012, Fermi \& Jansky Proceedings, arXiv:1204.6706 

\bibitem[Marscher(2013)]{ma2013}
Marscher, A. P., 2013, submitted to ApJ

\bibitem[Rybicki \& Lightman(1979)]{rl1979} 
Rybicki, G. B., \& Lightman, A. P., 1979, Radiative processes in
astrophysics, \\ John Wiley \& Sons, New York

\bibitem[Shakura \& Sunyaev(1973)]{ss1973}
Shakura, N. I., \& Sunyaev, R. A., 1973, A \& A, 24, 337
 
\bibitem[Sikora et al.(1994)]{sbr1994}
Sikora, M., Begelman, M. C., \& Rees, M. J., 1994, ApJ, 421, 153

\bibitem[Sikora et al.(1997)]{si1997}
Sikora, M., Madejski, G., Moderski, R., \& Poutanen, J., 1997, ApJ, 484, 108 

\bibitem[Sokolov \& Marscher(2005)]{sm2005}
Sokolov, A., \& Marscher, A. P., 2005, ApJ, 629, 52

\bibitem[Stern \& Poutanen(2011)]{sp2011}
Stern, B. E., \& Poutanen, J., 2011, MNRAS, 417, L11

\end{thebibliography}
\end{document}